\documentclass{aa}  
\usepackage{graphicx}
\usepackage{txfonts}
\usepackage{natbib}
\bibpunct{(}{)}{;}{a}{}{,}
\begin{document}
\title{New deep {\em XMM-Newton} observations \\ 
to the west of the $\sigma$~Orionis cluster} 
\titlerunning{New deep {\em XMM-Newton} observations to the west  of the $\sigma$~Orionis cluster}
%
%
\author{J. L\'opez-Santiago
    	\and
     	J. A. Caballero}
\offprints{Javier L\'opez-Santiago, \email{jls@astrax.fis.ucm.es}}
\institute{Departamento de Astrof\'{\i}sica y Ciencias de la Atm\'osfera, 
Facultad de Ciencias F\'{\i}sicas, Universidad Complutense de Madrid, 
E-28040 Madrid, Spain}
\date{Received 27 June 2008 / Accepted 02 September 2008}

\abstract
{} 
{The objective of this study is to determine the general X-ray properties
of the young stars in the external regions of the $\sigma$~Orionis cluster 
($\tau \sim 3$\,Ma, $d \sim 385$\,pc) and yield constraints on the X-ray 
emission of brown dwarfs.}
{We carried out a careful analysis of public data taken in an 
unexplored region to the west of the centre of the cluster with the three 
EPIC cameras onboard the \textit{XMM-Newton} mission. We looked for
new X-ray young stars among the 41 identified X-ray sources in the area 
with maximum likelihood parameters $L >$ 15 by cross-correlation with 
the USNO-B1, DENIS, and 2MASS databases.}
{Based on colour-colour, colour-magnitude, and hardness ratio diagrams, and 
previous spectroscopic, astrometric, and infrared-flux excess information, 
we classified the optical/near-infrared counterparts of the X-ray sources
into: young stars (15), field stars (4), galaxies (19), and sources of unknown
nature (3).
Most of the X-ray detections, including those of nine young stars, are new.
We derived the X-ray properties (e.g. temperatures, 
metallicities, column densities) of the twelve young stars 
with the largest signal-to-noise ratios. The X-ray parameters determined here
are in well agreement with those found in the cluster centre, where  
the stellar density is higher. 
There is no relation between infrared excess and column density from X-ray 
measurement in our data.
We detected flaring events in two young 
stars of the sample. One of them showed a very large ($\sim 30$) relative 
increase in flux. Both stars showed high coronal temperatures during the
observation. Finally, we determined upper limits to the flux of the young
stars and bright brown dwarfs not detected by our searching 
algorithm.}
{}
\maketitle
%

\section{Introduction}
\label{section.introduction}

\begin{figure*}
\centering
\includegraphics[width=0.48\textwidth]{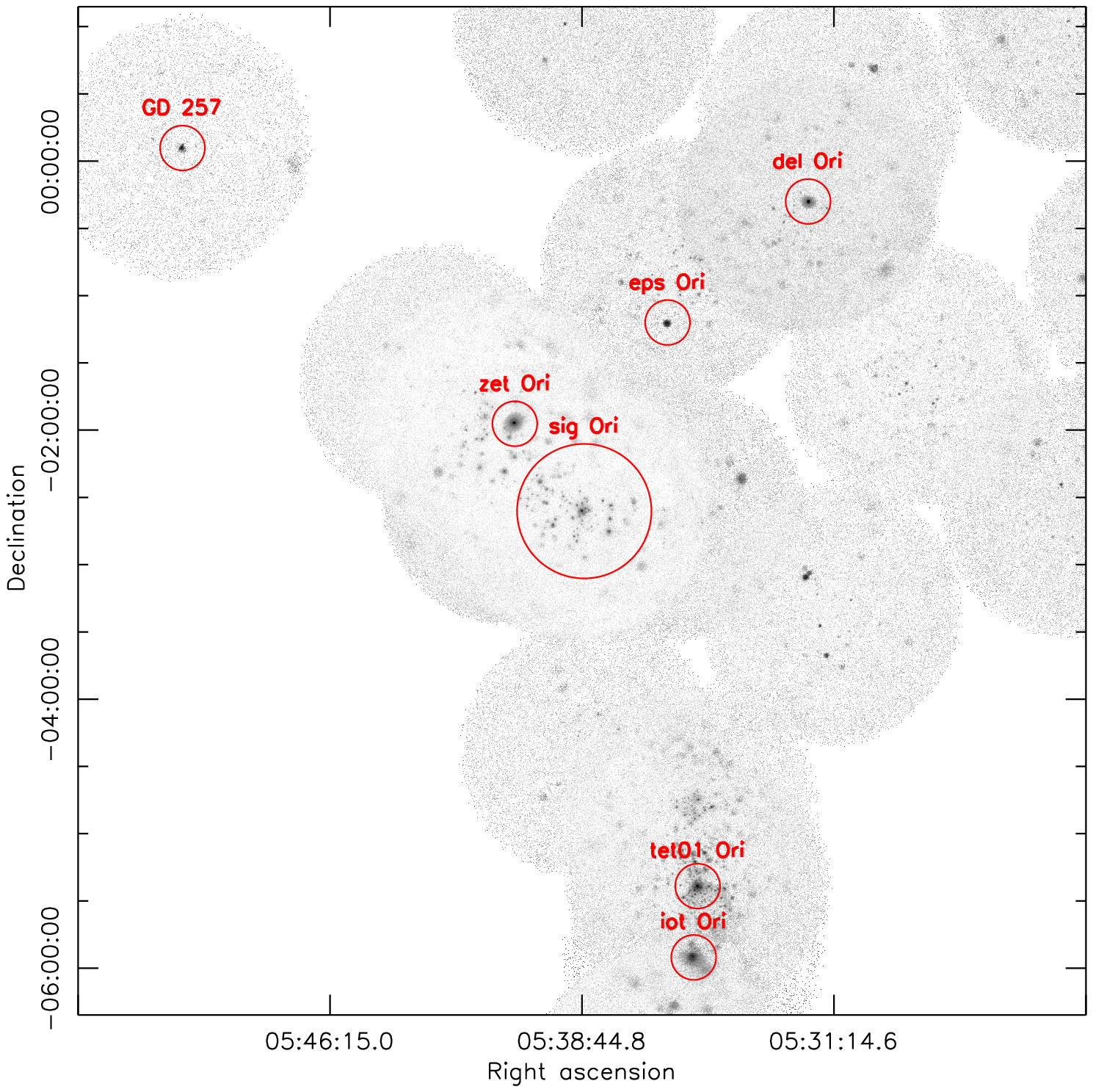}
\includegraphics[width=0.48\textwidth]{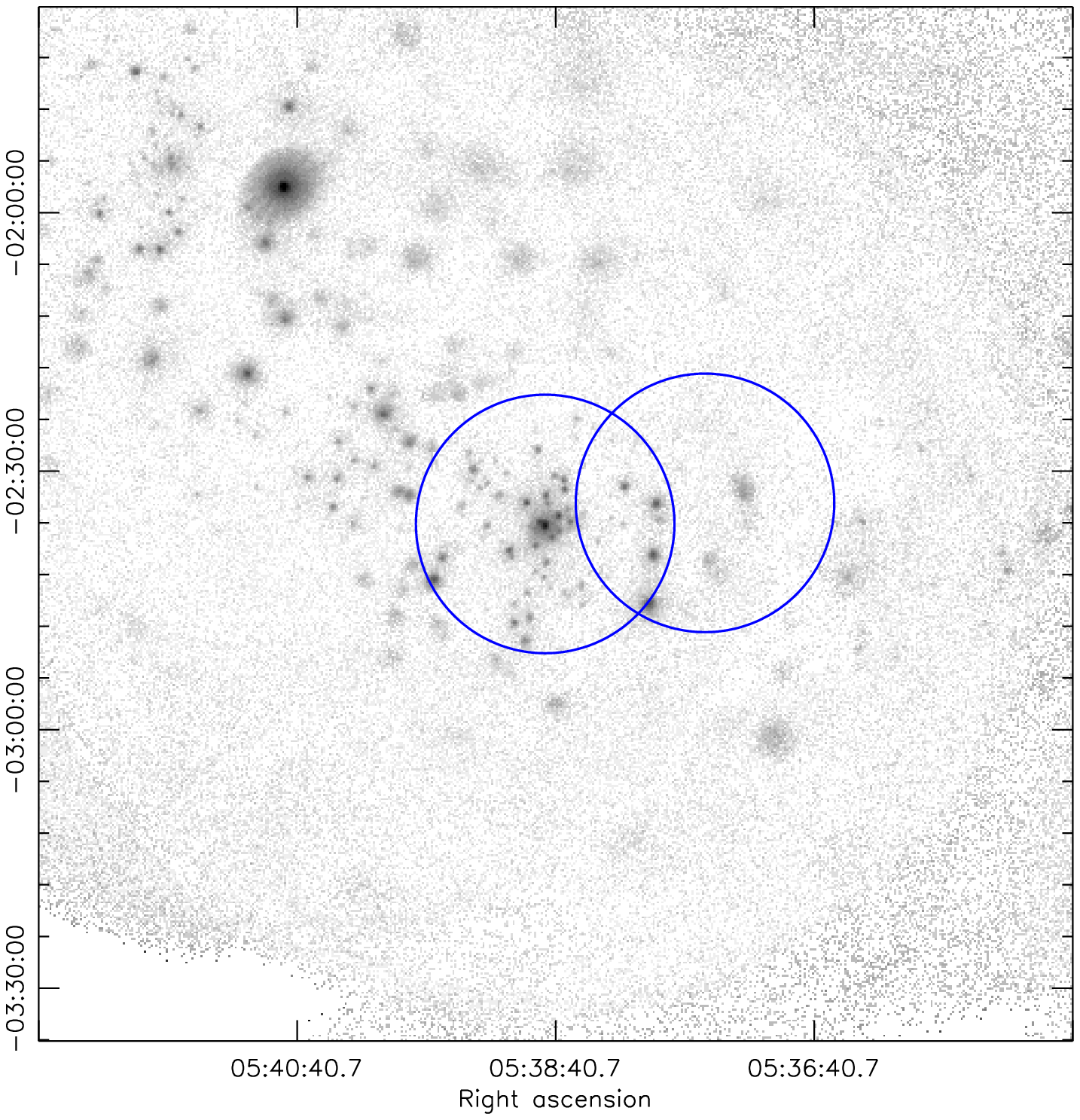}
\caption{Position Sensitive Proportional Counters (PSPC) 2.0\,deg cut-off, 
intensity-scaled, summed pointed observations with {\em ROSAT} 
(0.1--2.4\,keV) centred on the Trapezium-like system $\sigma$~Ori. 
North is up and east is left.
{\em Left:} the Orion Belt, to the centre and north, and the Orion Nebula
Cluster, to the south ($\sim$7$\times$7\,deg$^2$).
The brightest Orion complex stars in the area and the field white dwarf
\object{GD~257} are indicated and labelled. 
{\em Right:} zoom of the inner part ($\sim$2$\times$2\,deg$^2$).
Circles represent the two regions observed by \textit{XMM-Newton}:
the first is centred on $\sigma$~Ori and was analyzed by \citet{fra06};  the 
second is centred on S Ori 55 and is analyzed in this work.
Note the three X-ray sources of approximately the same brightness, aligned in
the north-south direction, at about 15\,arcmin to the west of $\sigma$~Ori.
Colour versions of all our figures are available in the electronic publication.
} 
\label{rosat}
% ...
\end{figure*}

The \object{$\sigma$~Orionis} cluster ($\tau \sim 3$\,Ma, $d \sim$ 385\,pc) is
a ``well-equipped laboratory'' for studying the formation, evolution, and
astrophysical properties of high-mass, solar-like, and low-mass stars, brown
dwarfs, and planetary-mass objects below the deuterium burning mass limit
\citep{wal74,gro82,zap00,bej01,ken05,cab07}.
The cluster takes the name from the Trapezium-like system $\sigma$~Ori in its
centre (we use the abridged name, $\sigma$~Ori, for the --at least-- sextuple
system, and the full name, $\sigma$~Orionis, for the eponymous cluster).
In spite of its early discovery \citep{gar67}, it was not until the pioneer
works by \citet{wol96} and \citet{wal97}, who detected an over-density of
X-ray emitters and pre-main sequence stars (see Fig.~\ref{rosat}), when the
$\sigma$~Orionis cluster started taking importance.
Afterwards, the detection of Herbig-Haro objects \citep{rei98}, brown
dwarfs \citep{bej99}, the least massive object directly detected out of
the Solar System \citep[\object{S\,Ori~70};][]{zap02b}, and
planetary-mass objects with discs \citep{zap07} have 
transformed $\sigma$~Orionis in a cornerstone for the study of stellar and
substellar populations in very young open clusters.
See \citet{cab07a,cab08c} 
for comprehensive $\sigma$~Orionis data~compilations.  

Regarding high energies in the cluster, the soft X-ray emission (0.2--3.5\,keV)
of \object{$\sigma$~Ori~A}F--B 
(O9.5V + B0.0V + B0.5V -- Caballero 2008b; D. M. Peterson et al. in prep.)
was first revealed by the {\em Einstein Observatory} \citep{chl89}.
The X-ray emission in such an early-type star is thought to come from
shock-heated gas in the radiatively driven stellar winds.
\citet{ber94} 
found no X-ray variability during a long-term
monitoring with the {\em R\"ontgensatellit} ({\em ROSAT}) of the triple
system (unresolved in their observations), which showed the stability of their
winds. Later, \citet{san04} and \citet{wal07}
presented high energy resolution spectra of $\sigma$~Ori~AF--B
obtained with the {\em X-ray Multi-Mirror Mission - Newton} ({\em
XMM-Newton}) and the {\em Chandra X-ray Observatory}, respectively. 
Very recently, \citet{ski08}, 
from data obtained with the High
Energy Transmission Grating Spectrometer onboard {\em Chandra}, considered
possible alternatives to the radiative wind shock, including
magnetically-confined and colliding wind shocks.

\citet{san04} reported a strong X-ray flare on the
magnetic B2Vp star \object{$\sigma$~Ori~E}, the second brightest star in
the cluster, confirming the previous discovery
of a flare on the star by \citet{gro04} with {\em ROSAT} data.
Caballero et~al. (in~prep.) have also detected another flare using
alternative {\em ROSAT} data.
However, X-ray flares are not expected in the wind scenario of early-type stars.
The low-mass companion hypothesis supported by \citet{san04} 
and discussed in the literature since the early 1970s \citep[see][]{lan78}
has been lately confirmed by \citet{bou08} 
using a multi-conjugate adaptive optics system ($\sigma$~Ori~E and its fainter
companion are separated by only 0.330\,arcsec). 

In contrast to early-type stars, the origin of the X-ray emission in
low-mass stars resides on the 
high temperatures of their coronae. 
The first detailed analysis of young, low-mass X-ray stars in the
$\sigma$~Orionis cluster was carried out by \citet{wol96} 
using {\em ROSAT}. His work was surpassed, however, by the exhaustive 
analysis of the full EPIC/{\em XMM-Newton} field centred on $\sigma$~Ori~AF--B
by \citet{fra06}.
They detected 175 X-ray sources, 88 of which they identified with
$\sigma$~Orionis cluster members and candidate members.
\citet{fra06} 
found no significant difference in the spectral
properties of 23 classical and weak-line T~Tauri stars classified in the
literature.  
The formers, however, tended to be less X-ray luminous than the last.
This trend was also found in the cluster centre by \citet{cab07b} 
using {\em Chandra} data, which confirmed, for instance, the earlier 
results by \citet{neu95} 
in Taurus. Additional intermediate- and low-mass stars with X-ray emission found with 
{\em Einstein}, the {\em Advanced Satellite for Cosmology and Astrophysics} ({\em
ASCA}), {\em ROSAT}, and {\em Chandra} have been tabulated by 
\citet{alc96}, \citet{nak99}, \citet{ada04}, \citet{cab07a,cab08c}, and 
\citet{ski08}.
The X-ray variability of 23 young stars (and one galaxy) with {\em ROSAT} has
been recently investigated by Caballero et~al. (in~prep.).

The X-ray activity extends down to below the hydrogen burning mass limit in
$\sigma$~Orionis. 
\citet{mok02} firstly investigated the X-ray emission near the
substellar limit of the cluster.
Of the seven very low-mass $\sigma$~Orionis members in the error box of a {\em
ROSAT} source, three of them were classified as unambiguous detections.
However, two of the actual emitters are very close ($\rho \sim$ 4--5\,arcsec)
to young low-mass cluster stars, that are probably the actual X-ray sources and
not the brown dwarfs \citep{cab06,cab07}.
The third X-ray brown dwarf candidate in \citet{mok02}, 
\object{Mayrit~633059} (S\,Ori~3), is $\Delta J \sim$ 1.2\,mag brighter than the
stellar/substellar boundary (i.e. it is a~star).

To date, there remain only three {\em confirmed} $\sigma$~Orionis brown dwarfs
with X-ray emission \citep{fra06}: 
\object{Mayrit~433123} (S\,Ori~25), \object{Mayrit~487350} ([SE2004]~70), and
\object{Mayrit~396273} (S\,Ori~J053818.2--023539).
The spectra of the three of them have Li~{\sc i} $\lambda$6707.8\,{\AA} in
absorption and/or low gravity features (e.g. abnormal alkali strength).
Mayrit~433123 is a well known T~Tauri substellar analog with H$\alpha$ in
strong emission, as well as He~{\sc i} \citep{bej99,bar03,muz03}.
It is, besides, one of the very few brown dwarfs whose rotational velocity
corrected from inclination has been derived 
\citep[$v$ = 14\,km\,s$^{-1}$;][]{cab04}.
Mayrit~487350 underwent a flare during \citet{fra06}'s 
observations,
and might be the primary of the widest exoplanetary system detected so far 
\citep[$r \sim$ 1700\,AU;][]{cab06b}.
Finally, Mayrit~396273 (the faintest brown dwarf of the trio) has been poorly
investigated \citep{ken05,max08}. 

With the aim of exploring the X-ray emission of extremely low-mass
$\sigma$~Orionis members, F.~Mokler conducted new deep {\em XMM-Newton}
observations to the west of the cluster, complementing 
\citet{san04} and \citet{fra06}'s observations, that imaged the cluster centre.
The new observations were centred, in its turn, on Mayrit~1158274
(\object{S\,Ori~55}), a substellar object at the brown dwarf/planet boundary
with strong variable H$\alpha$ emission and infrared flux excess longwards
of 5\,$\mu$m \citep{zap02a,zap07,luh08}. 


We use the deep \textit{XMM-Newton} observations to the west of the
 $\sigma$~Orionis cluster to determine general X-ray properties
of the young stars in the cluster external region, where the stellar density 
drops. To complete our work, we give upper limits to the X-ray emission 
of the brown dwarfs in the field. Our intention is to determine
the properties of the cluster X-ray emitters down to the substellar
boundary. Once identified the X-ray stellar sources in the 
field, we conduct a spectral analysis to derive their basic parameters
(e.g. energies of the thermal components, coronal metallicities,
hydrogen column densities). We also look for flares during the
X-ray monitoring and long-term variability. Finally, we complement
our results with data from the literature to study the relation
between X-ray emission and red optical-near infrared colours,
which are indicative of spectral type and of the presence of surrounding
discs.

\section{Observation and reduction}
\label{section.observations}

   \begin{table}
      \caption[]{Basic data of the observations and reduction.}  
         \label{table.observations}
     $$ 
         \begin{tabular}{ll}
            \hline
            \hline
            \noalign{\smallskip}
			& Observations \\  
            \noalign{\smallskip}
            \hline
            \noalign{\smallskip}
Telescope		& {\em XMM-Newton} \\  
Instrument		& EPIC \\  
Cameras			& PN+MOS1+MOS2 \\  
Mode			& Full frame imaging \\  
Filter			& Thin \\  
Centre $\alpha$ (J2000)	& 05 37 31.4 \\
Centre $\delta$ (J2000)	& $-$02 33 41 \\
Start observations (UT) & 2003 08 30, 13:52 \\  
End observations (UT) 	& 2003 08 31, 09:02 \\  
PN exposure time 	& 42\,ks \\  
MOS1 exposure time 	& 50\,ks \\  
MOS2 exposure time 	& 52\,ks \\  
Total duration 		& 69\,ks \\  
           \noalign{\smallskip}
            \hline
         \end{tabular}
     $$ 
   \end{table}
%

This {\em XMM-Newton} observation of the western region of the $\sigma$\,Orionis
cluster (ID 0148300101) was performed in the revolution 682, between 2003 
August 30 and 31, for a total duration of 69\,ks. 
The three European Photon Imaging Cameras (EPIC) were operated simultaneously in
full frame mode. 
In Table~\ref{table.observations}, we give the most important details of the
observation. 

Data reductions followed the standard procedures. We used the version 7.1.0
of the {\em XMM-Newton} Science Analysis System software (SAS) to derive a table
of calibrated events. 
Then, we applied different filters to eliminate bad events and noise. 
Finally, high flaring background periods were removed in each detector
separately to maximize the signal-to-noise ratio of weak sources. 
Although the observation was strongly affected by high background, the final
effective exposure times (42, 50, and 52\,ks for the PN, MOS1, and MOS2
detectors, respectively) are still longer than the observation analysed by
\citet{fra06}. 
Nevertheless, both observations are complementary since they slightly
overlap at the eastern edge of our survey~area. 

\begin{figure}
\centering
\includegraphics[width=0.49\textwidth]{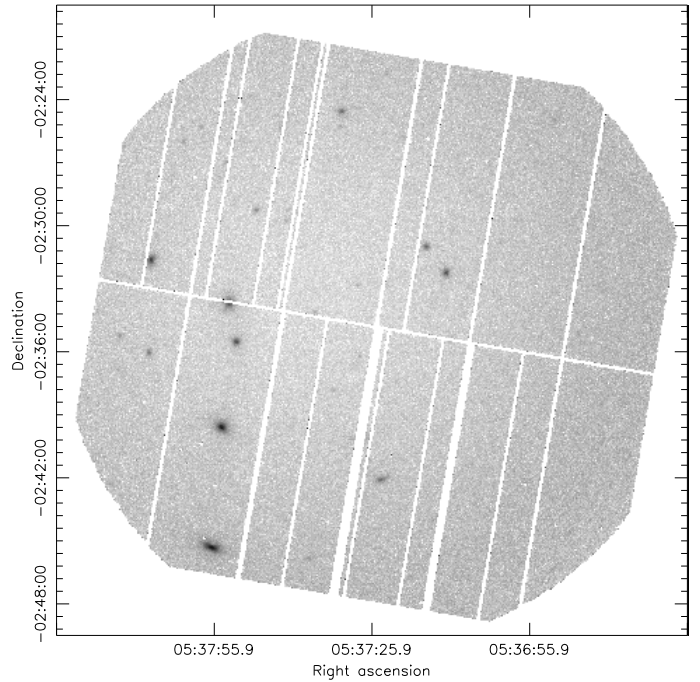}
\caption{Combined EPIC image of the western region of the
$\sigma$\,Orionis cluster in the 0.3--7.5\,keV energy band.
The three aligned X-ray sources in right window in Fig.~\ref{rosat} are now
splitted into four ones (the northernmost source falls just on a gap between
detectors).} 
\label{mosaic}
\end{figure}

In Fig.~\ref{mosaic}, we show the combined (filtered and calibrated) EPIC 
(PN+MOS1+MOS2) image. 
We created images in the chosen energy band (0.3--7.5\,keV) for each detector
individually.  
Then, we constructed exposure maps using the EEXMAP routine in the SAS software.
These maps contain the spatial efficiency of the instruments. 
By dividing the images by the exposure maps, we corrected them for the quantum
efficiency, filter  transmission, and mirror vignetting. 
The resultant images were divided by the corresponding effective area to account
for the difference in efficiency of the EPIC-PN and EPIC-MOS detectors. 
Finally, the task EMOSAIC was used to construct a combined PN+MOS image.
A false-colour image combining X-ray, optical, and near-infrared data is
shown in Fig.~\ref{falsecolour}.

\section{Analysis and results}
\label{section.analysis}

\subsection{X-ray source detection}
\label{section.identification}

Standard packages available in SAS were used to reveal the sources in 
this region. 
In particular, we made use of the multi-task EMLDETECT, that first performs a
\textit{sliding cell detection} and then uses a \textit{maximum likelihood
method} to derive parameters for each source.
The sliding cell method consists in a simple sliding-window-box detection
algorithm.
We used a low signal-to-noise ratio above the background ($\sim 4 \sigma$) to
allow the detection of as many sources as possible. 
The list of sources is given, in its turn, as an input in the maximum likelihood
method.
Final derived parameters of EMLDETECT for each source are: 
total source count-rate ($CR$), 
likelihood of detection ($L$), 
and hardness ratio ($HR$). 
The reader is referred to the specific SAS documentation 
page\footnote{\tt http://xmm.vilspa.esa.es/sas/7.1.0/doc/emldetect/index.html.} 
for more details on the procedure.

We performed the source detection for each camera dataset in three different
energy bands:  
soft (0.3--2.0\,keV), medium (2.0--4.5\,keV), and hard (4.5--7.5\,keV).  
Cleaned event-file corrected images (from bad pixels and noise) were used
at this stage (see Section~\ref{section.observations}).
We ignored the energy channels below 0.3\,keV, because of the strong noise
affecting the EPIC detectors under this value, and those above 7.5\,keV, since 
only background is present at high energies. 
The energy bands selected here are very similar to those used in the {\em
XMM-Newton} Bright Serendipitous Survey 
\citep[XBSS;][]{cec04}\footnote{The $HR_2$ and $HR_3$ hardness ratios in 
\citet{cec04} are identical to our $HR_1$ and $HR_2$ ones, except for our soft 
band extending down to 0.3\,keV instead of 0.5\,keV.}. 
The hardness ratios constructed using these bands,

\begin{eqnarray}
HR_{1} & = & \frac{CR_{\rm medium} - CR_{\rm soft}}
                  {CR_{\rm medium} + CR_{\rm soft}} \\
HR_{2} & = & \frac{CR_{\rm hard} - CR_{\rm medium}}
                  {CR_{\rm hard} + CR_{\rm medium}},
\end{eqnarray}

\noindent showed up as a very powerful tool to separate stellar sources from
extragalactic ones.  
Non-absorbed coronal sources show, in general, low hardness-ratios ($HR_1
\lesssim$ --0.75), while active galactic nuclei (AGNs) of galaxies present
higher values because of both their high intrinsic absorption and emission
mechanisms.  

\begin{figure}
\centering
\caption{False-colour composite images centred on four representative X-ray
sources (indicated with a 10\,arcsec-radius circle): NX~5 
({\em top left}), NX~11 ({\em top right}), NX~30 ({\em bottom left}), and 
NX~32 ({\em bottom right}). DSS-2 plates at photographic $B_J R_F I_N$ bands 
are for blue, green, and red colours, respectively.
Size is about 1.5$\times$1.5\,arcmin$^2$.
North is up, east is left.
In the NX~5 (OriNTT~429~AB) field, note the photometric cluster member candidate
Mayrit~1416280 \citep[OriNTT~429~C;][]{cab08c}
at $\rho \sim$ 12.1\,arcsec to the north of the spectroscopic binary. 
In the NX~11 field, the 2MASS and DENIS photo-centroids of the two objects
discussed in the text are at $\sim$4\,arcsec to the X-ray source.
NX~30 (Mayrit~783254) is a young, early K-type dwarf with lithium in
absorption. NX~32 (2E~1456) is a galaxy with a strong X-ray emission
(Section~\ref{section.galaxies}).
\textbf{Note: images are available only in the A\&A version.}}  
\label{figure.NXBRI}
\end{figure}

After a careful visual inspection of the sources in the initial list, which 
allowed us to reject artifacts (bad pixels in the gaps between chips and
near the borders, multiple detections, and cosmic rays), we kept 41 X-ray
sources with a maximum likelihood parameter $L \ge 15$.  
This parameter is defined as the Neperian logarithm of the probability $P$ that
the observed counts come from random Poissonian fluctuations ($L = - \ln P$).
Therefore, we kept those sources with a probability $P \lesssim 3.0 \times 10^{-7}$
of being spurious (non-Poissonian effects --e.g. unidentified artifacts-- may
affect this probability when $L \sim 15$). 

The final list of X-ray sources is given in Table~\ref{table.sources}. 
For each source we give its position in equatorial coordinates, maximum 
likelihood of detection, count-rate obtained in each detector in the merged
energy band 0.3--7.5\,keV, and the hardness ratios ($HR_1$ and $HR_2$).
Several sources fell in CCD gaps or in non-overlapping fields of view of the
three cameras, and lack, as a result, some measurements.
Six X-ray sources (NX~4, 8, 16, 25, 31, and 35) are detected only in one camera,
and with maximum likelihood parameters in the interval $15 \le L \le 24$, which
calls its truly emission into question.

\subsection{Cross-identification and source classification}
\label{section.cross}

\begin{figure}
\centering
\includegraphics[width=0.49\textwidth]{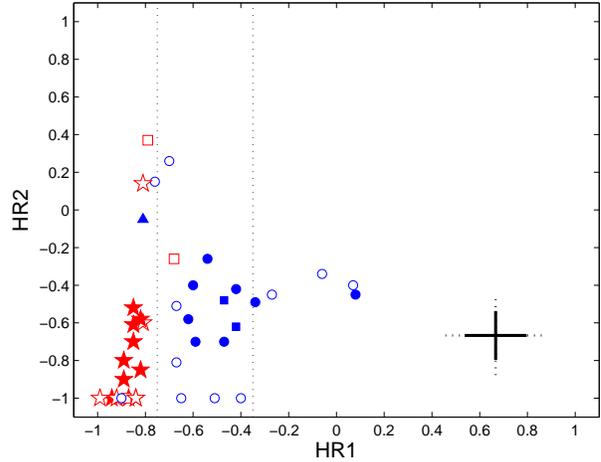}
\caption{Hardness ratios diagram \citep[$HR_2$ vs. $HR_1$; compare with fig.~5 
in][]{cec04}. The different marker symbols represent:
confirmed young stars in the Mayrit catalogue (--red-- filled star symbols),
candidate young stars (--red-- open star symbols), 
field stars (--red-- open squares), 
galaxies with near infrared and/or optical counterpart (--blue--
filled circles), 
galaxies without near infrared or optical counterpart (--blue-- open circles), 
NX~11 (--blue-- filled up-triangle), 
and NX~23 and~41 (--blue-- filled squares). 
The dotted lines at constant $HR_1$ correspond to the locus enclosing
$\sim$90\,\% of type-1 AGNs in the {\em XMM-Newton} bright serendipitous
survey sample \citep{cec04}.
In the lower right corner we report the median (solid) and the 90\,\% (dashed)
percentile of the $HR_1$ and $HR_2$ errors.}  
\label{diagram.HR1HR2}
\end{figure}

First, we looked for optical and near-infrared counterparts of the X-ray
sources in the USNO-B1, DENIS, and 2MASS catalogues \citep{epc97,mon03,skr06}
using Aladin \citep{bon00}. 
The maximum positional offset allowed for an object to be declared a
{\em possible} counterpart was fixed at a conservative value of 10\,arcsec, 
bearing in mind the point-spread-function (PSF) of the EPIC cameras and their 
astromeric accuracy.
For these cameras, the 80\% of the PSF is contained in a circle with
a diameter of 15\,arcsec. For additional values of angular separations between
X-ray and near-infrared counterparts in $\sigma$~Orionis, see \citet{fra06}.
The mean offset between {\em XMM-Newton} and 2MASS/USNO-B1 positions 
for the 30 X-ray sources with near-infrared/optical counterparts and their 
standard deviations are $\Delta \alpha = -2.2 \pm 1.8$\,arcsec, 
$\Delta\delta = -1.7 \pm 1.4$\,arcsec (i.e. the X-ray sources are 
situated, on average, at 1.2$\sigma$ to the southwest of the 
near-infrared/optical 
counterparts). The non-zero offset is, however, very small and does not 
affect our results. The majority of the 
near-infrared/optical counterparts of our sources are inside a radius of
5 arcsec. 
This value has been typically used as a good estimation of the effective 
$\sim 90$\% confidence radius of uncorrected positions \citep[e.g.][]{wat03}. 
Note that the statistical error-circle for a faint XMM-Newton source 
has $\sigma_\mathrm{stat} =$ 1--2\,arcsec.

The results of our cross-correlation are provided in 
Table~\ref{table.counterparts}. We compiled equatorial coordinates, 
$B_J$ (USNO-B1), $i$ (DENIS), and $JHK_{\rm s}$ (2MASS) magnitudes (or lower 
limits), and alternative names of the 41 X-ray sources. 
Coordinates were taken from Tycho-2, 2MASS, or USNO-B1 (with this order of
preference).
For the three brightest stars in the sample, we give Tycho-2 $B_T$ magnitudes
\citep{hog00} 
instead of photographic $B_J$, which is the average of the
two USNO-B1 measurements (``$B1$'' and ``$B2$'').
Uncertainties of the $B_J$ and $i$ magnitudes are $\sim$\,0.5 and 
0.2--0.3\,mag in this particular area.
See also \citet{cab08c}
for a discussion on the validity of
the DENIS $i$-band magnitude when $\delta i$ = 1.00\,mag.
Fig.~\ref{figure.NXBRI} shows the optical counterparts of four
representative X-ray sources and illustrates the cross-identification.

Up to eleven X-ray sources have no optical or near infrared counterpart at less
than 12\,arcsec; 
the nearest optical neighbours to some of these X-ray sources are located at
more than 30\,arcsec (40\,arcsec for NX~25).
For them, we provide the lower limits of their magnitudes from the completeness
magnitudes of each survey.
Besides, five X-ray sources have no 2MASS counterpart at 1.2--2.2\,$\mu$m, but
are catalogued by USNO-B1.
Of them, only one source, NX~12, was also identified by DENIS at the $i$ band.
We list $I_N$ instead of $i$ in Table~\ref{table.counterparts} for sources NX~3
and NX~18, that were found at the three photographic bands ($B_J R_F I_N$).
The remaining 25 X-ray sources have USNO-B1, DENIS, {\em and} 2MASS counterparts
at less than $\sim$\,6 arcsec.
Actually, the source NX~11 has {\em two} possible optical/near infrared
counterparts (see Fig.~\ref{figure.NXBRI}); it will be discussed~below. 

\begin{figure*}
\centering
\includegraphics[width=0.49\textwidth]{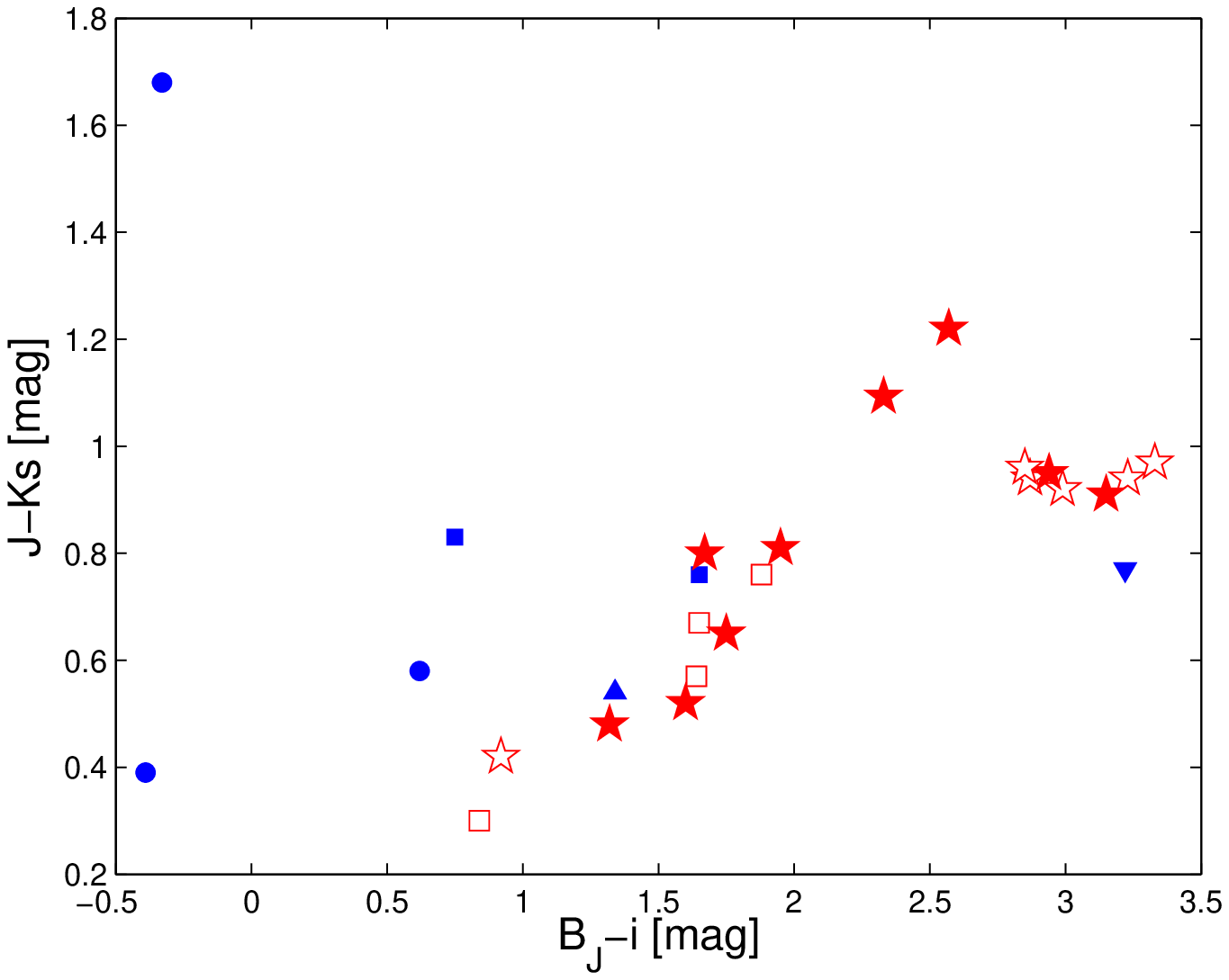}
\includegraphics[width=0.49\textwidth]{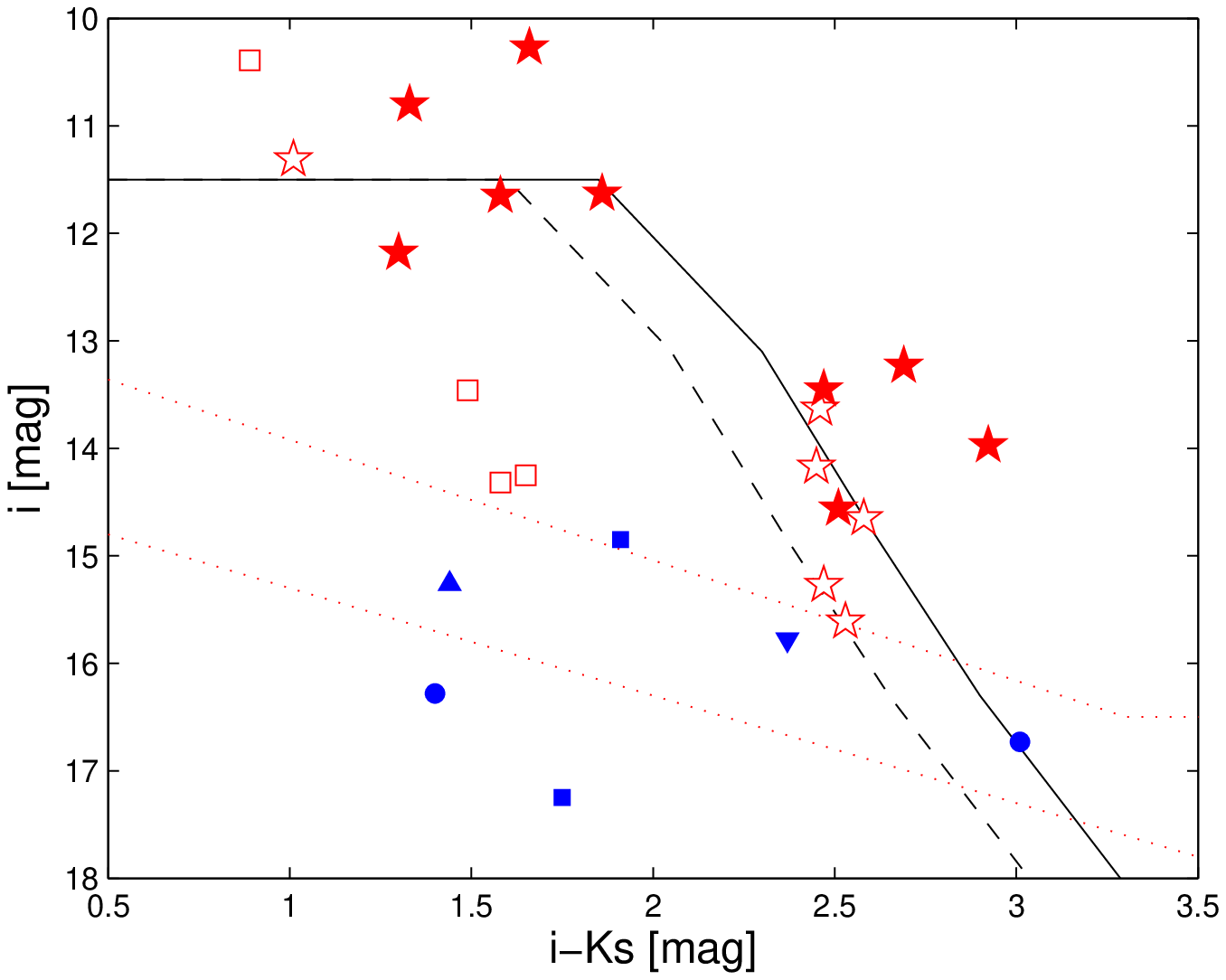}
\caption{Colour-colour and colour-magnitude diagrams.
The marker symbols are as in Fig.~\ref{diagram.HR1HR2}.
The (blue) filled up- and down-triangles are for the brighter and fainter
(in $B_J$) counterparts of NX~11, respectively.
{\em Left:} $J-K_{\rm s}$ vs. $B_J - i$ colour-colour diagram.
The star sequence runs from ($B_J - i$, $J-K_{\rm s}$) $\sim$ (1.0, 0.4)\,mag to
$\sim$ (2.5, 1.0)\,mag.
At redder $B_J - i$ colours, the $J-K_{\rm s}$ colours keep roughly constant at
$\sim$ 1.0\,mag except for disc harbour stars, like the two class~II stars
at ($B_J - i$, $J-K_{\rm s}$) $\sim$ (2.3--2.6, 1.10--1.25)\,mag.
{\em Right:} $i$ vs. $i-K_{\rm s}$ colour-magnitude diagram.
The dotted (red) lines are approximate completeness and detection limits of the
combined DENIS/2MASS cross-correlation.
The solid (black) line is the criterion for selecting candidate cluster stars
and brown dwarfs without known features of youth in $\sigma$~Orionis used by
\citet{cab08c}. 
The dashed (black) line is the criterion shifted bluewards by 0.25\,mag.}  
\label{diagram.CCandCM}
\end{figure*}

We have classified the 41 X-ray sources and their corresponding optical/near
infrared counterparts into young stars (15), field stars (4), galaxies (19), and
sources of unknown nature (3) depending on their positions in the hardness ratio
(Fig.~\ref{diagram.HR1HR2}), colour-colour, and colour magnitude
(Fig.~\ref{diagram.CCandCM}) diagrams, and on spectroscopic, astrometric,
infrared photometric, and point spread function (PSF) information.
Table~\ref{table.class} summarizes our classification.
The last column indicates the Mayrit number for $\sigma$~Orionis cluster members
and candidates \citep{cab08c}.

   \begin{table*}
      \caption[]{Classification of X-ray sources.}  
         \label{table.class}
     $$ 
         \begin{tabular}{c c l l l l}
            \hline
            \hline
            \noalign{\smallskip}
NX	& [FPS2006]$^{a}$	& Remarks	& Reference$^{b}$	& Class			& Mayrit 	\\
            \noalign{\smallskip}
            \hline
            \noalign{\smallskip}
1	& ...	& No optical/nIR counterpart		& ...			& Galaxy		& ... 		\\
2	& ...	& No optical/nIR counterpart		& ...			& Galaxy		& ... 		\\
3	& ...	& Only optical counterpart		& ...			& Galaxy		& ... 		\\
4	& ...	& Phot. young candidate 		& Ca07a			& Young star candidate 	& 1456284 	\\
5	& ...	& Li~{\sc i}, H$\alpha$, SB2		& Lee94, Ca06		& Young star 		& 1415279AB 	\\
6	& ...	& Blue $i-K_{\rm s}$			& ...			& Field star 		& ...	 	\\
7	& ...	& Li~{\sc i}				& Ca06			& Young star 		& 1374283 	\\
8	& ...	& No optical/nIR counterpart		& ...			& Galaxy		& ... 		\\
9	& ...	& Blue $i-K_{\rm s}$			& ...			& Field star 		& ...	 	\\
10	& ...	& Extended PSF				& ...			& Galaxy		& ... 		\\
11	& ...	& Extended PSF/binary system		& ...			& Unknown 		& ... 		\\
12	& ...	& Extended PSF, only optical counterpart& ...			& Galaxy		& ... 		\\
13	& ...	& Extended PSF, blue/red, strong X-ray  & ...			& Galaxy 		& ... 		\\
14	& ...	& New phot. young candidate		& ...			& Young star candidate	& 1149270	\\
15	& ...	& No optical/nIR counterpart		& ...			& Galaxy		& ...		\\
16	& ...	& New phot. young candidate		& ...			& Young star candidate	& 1344302	\\
17	& ...	& Phot. young candidate, Li~{\sc i}?, H$\alpha$? & Sh04, Ca06	& Young star candidate	& 1298302	\\
18	& ...	& Only optical counterpart		& ...			& Galaxy		& ...		\\
19	& ...	& Only optical counterpart		& ...			& Galaxy		& ...		\\
20	& ...	& Li~{\sc i}				& Sh04, Ca06		& Young star		& 1027277	\\
21	& ...	& New phot. young candidate		& ...			& Young star candidate	& 1160240	\\
22	& ...	& No $\mu$				& Ca07a 		& Field star		& ...		\\
23	& ...	& Blue $i-K_{\rm s}$			& ...			& Unknown		& ...		\\
24	& ...	& Blue $i-K_{\rm s}$			& ...			& Field star		& ...		\\
25	& ...	& No optical/nIR counterpart		& ...			& Galaxy		& ...		\\
26	& ...	& No optical/nIR counterpart		& ...			& Galaxy		& ...		\\
27	& 1	& Li~{\sc i}, H$\alpha$, M1-3		& Wo06, Ca08c		& Young star		& 797272	\\
28	& 2	& Li~{\sc i}, broad H$\alpha$, K0	& Wo96, Ca06		& Young star		& 789281	\\
29	& ...	& No optical/nIR counterpart		& ...			& Galaxy		& ...		\\ 
30	& 3	& Li~{\sc i}, H$\alpha$, K0		& Wo96  		& Young star		& 783254	\\ 
31	& ...	& No optical/nIR counterpart		& ...			& Galaxy		& ...		\\
32	& ...	& Extended PSF, blue/red, strong X-ray  & Ca08d 		& Galaxy		& ...		\\
33	& ...	& Only optical counterpart		& ...			& Galaxy		& ...		\\
34	& ...	& Phot. young candidate 		& SE04, Sh04		& Young star candidate	& 887313	\\
35	& 5	& No optical/nIR counterpart		& ...			& Galaxy		& ...		\\
36	& 7	& Strong H$\alpha$, class II		& Wi89, Sh04, He07	& Young star		& 662301	\\
37	& 8	& Phot. young candidate, X-ray  	& Wo96, Fr06		& Young star candidate	& 615296	\\
38	& 9	& Strong H$\alpha$, class II		& Wi91, Sh04, He07	& Young star		& 547270	\\
39	& 12	& No optical/nIR counterpart		& ...			& Galaxy		& ...		\\
40	& ...	& No optical/nIR counterpart		& ...			& Galaxy		& ...		\\
41	& 14	& Blue $i-K_{\rm s}$ 			& ...			& Unknown 		& ... 		\\ 
            \noalign{\smallskip}
            \hline
         \end{tabular}
     $$ 
\begin{list}{}{}
\item[$^{a}$] X-ray source number in \citet{fra06}. 
\item[$^{b}$] Reference abbreviations --
Wi89: \citet{wir89}; 
Wi91: \citet{wir91}; 
Lee94: \citet{lee94}; 
Wo96: \citet{wol96}; 
SE04: \citet{sch04}; 
Sh04: \citet{she04}; 
Fr06: \citet{fra06}; 
Ca06: \citet{cab06}; 
Ca07a: \citet{cab07a}; 
He07: \citet{her07}; 
Ca08c: \citet{cab08c}; 
Ca08d: \citet{cab08d}. 
\end{list}
   \end{table*}

\subsubsection{Young stars}
\label{section.youngstars}

This class comprises the optical/near infrared counterparts of stars with known
features of youth or that follow the spectro-photometric sequence of the
$\sigma$~Orionis cluster.
We have splitted the 15 young stars, in its turn, into confirmed (8) and
candidate young stars (7).
Six of the eight confirmed young stars have Li~{\sc i} $\lambda$6707.8\,\AA~in
absorption in spectra taken by \citet{lee94}, \citet{wol96}, and \citet{cab06}:
NX~5, 7, 20, 27, 28, and 30 (NX~5, 27, and 28 have redundant detections).
There is information on its H$\alpha$ $\lambda$6562.8\,\AA~line for the six
stars.
In all cases, the H$\alpha$ emission seems to be chromospheric. 
Likewise, other two stars (NX~36 and~38) have both intense H$\alpha$ emission
detected with prism-objective and Schmidt plates \citep{wir89,wir91}
and infrared flux excess at the IRAC/{\em Spitzer} bands \citep{her07}
These effects are ascribed to the presence of circumstellar discs from where the
central object accretes material.
The two objects have been classified as classical T~Tauri stars (class~II).

There remain seven young star candidates.
Four of them were tabulated as photometric cluster member candidates by 
\citet{wol96}, \citet{sch04}, \citet{she04}, and \citet{cab06,cab07a}: 
NX~4, 17, 34, and~37.
The last star appears in catalogues prepared with {\em Einstein} data 
\citep{hab94}, was classified as a pre-main sequence {\em ROSAT} X-ray emitter
star by \citet{wol96}, and had a large unabsorbed X-ray luminosity ($L_{\rm X}$ =
30.66 [cgs]) in the {\em XMM-Newton} observations by \citet{fra06}, 
that make NX~37 to be a truly young active star.
\citet{cab06} 
obtained a poor optical spectrum of NX~17 where the Li~{\sc i}
and H$\alpha$ lines were hardly discernible.
The three remaining photometric candidates (NX~14, 16, and~21) are presented
here for the first time.
The three of them also follow the sequence of the young stars in the hardness
ratio, colour-colour, and colour-magnitude diagrams.
In reallity, they are slightly bluewards of the conservative selection criterion
in the $i$ vs. $i-K_{\rm s}$ diagram in fig.~4 in \citet{cab08c}, 
but still
redwards of the lower envelope of confirmed young stars in $\sigma$~Orionis.
As already mentioned by \citet{cab06b}, 
cluster members with X-ray
emission tend to have bluer optical/near-infrared colours than T~Tauri analogs.
Since most of the photometric searches in the cluster have been based on
colour-magnitude diagrams and, therefore, biased towards the detection of 
(red) T~Tauri analogs, the bluest part of the spectro-photometric cluster
sequence is not well defined yet and many X-ray stars still wait identification.
All the 15 young stars and candidates have hardness ratios $HR_1 <$ --0.75.

\subsubsection{Field stars}

The four field star candidates have hardness ratios consistent with their
probable stellar nature if uncertainties are taken into account.  
The brightest counterpart (NX~21) is a Tycho-2 star with accurate astrometry
that was classified as a non-cluster member by \citet{cab07a} 
based on its proper motion larger than 10\,mas\,a$^{-1}$.
Their colours match with those of F-type (interacting binary?) dwarfs. 
It is the only field star whose maximum likelihood parameter $L$ is larger
than~25.   

The remaining three field star candidates, that are among the faintest X-ray
sources in our sample, are presented here for the first time. 
They have very blue $i-K_{\rm s}$ colours for its $i$ magnitude if compared with
the $\sigma$~Orionis sequence in the colour-magnitude diagram (by up to
0.75--1.00\,mag), but follow the dwarf sequence in the colour-colour
diagram (Fig.~\ref{diagram.CCandCM}). 
They seem to be late K-type dwarfs in the foreground of $\sigma$~Orionis.

\subsubsection{Galaxies}
\label{section.galaxies}

The eleven X-ray sources without optical or near infrared counterpart are
probably background galaxies.
First, they have magnitudes $J \gtrsim$ 17.1\,mag, that correspond to the lower
mass part of the substellar domain in $\sigma$~Orionis \citep{cab07}. 
If cluster members, they would have magnitudes $i \gtrsim$ 21\,mag, which
would explain their null detection in the optical.
However, even in the most favourable case of an $\alpha$ index of the mass
spectrum ($\Delta N / \Delta M \propto M^{-\alpha}$) close to the Salpeter one
($\alpha$ = --2.35), which is in clear contradiction with recent determinations
of the initial mass function in the cluster \citep{cab07}, 
we would not expect such a large population of faint cluster members.
Secondly, these hypothetical faint cluster members would have extraordinary
large $L_{\rm X} / L_{\rm bol}$ ratios: the faintest X-ray brown dwarfs in
$\sigma$~Orionis (Section~\ref{section.introduction}) are brighter than $J$ =
15.5\,mag.
Thirdly, the great majority of the X-ray sources have hardness ratios $HR_1 >$
--0.75, more typical of extragalactic sources.
And finally, the contamination by background galaxies in $\sigma$~Orionis, some
of which may be X-ray emitters, gets more important at the faintest magnitudes.
See compilations of extragalactic sources in the Orion Belt (some of which were
previously identified as active stars) in \citet{cab08c}, \citet{cab08},
and \citet{cab08d}. 
As a result, the abundance of X-ray sources with faint optical counterparts or
without counterparts at all \citep[see Table B.1 in][]{fra06} is
naturally explained by a background population of galaxies 
\citep[especially, AGNs powered by massive black holes --][]
{ree84,sto91,com95,ale03,ued03}.

The extragalactic hypothesis is supported by the identification of four X-ray
sources in our field with extended PSFs in the optical and/or near-infrared 
(NX~10, 12, 13, and 32) and four USNO-B1 sources without near infrared 
counterpart.
The last sources have colours $B_J - J \lesssim$ 2.7\,mag, i.e. bluer than
late K-type stars, approximately.
Such blue colours at the corresponding faint magnitudes are inconsistent with
those of early-type stars at very long heliocentric distances 
\citep[at several Galactic height scales;][]{che01} and with late-type dwarfs
in the foreground of the cluster.
The low extinction towards $\sigma$~Orionis ($A_V \sim$ 0.3\,mag) cannot be
responsible of the faintness of these sources in the near infrared.

Two of the galaxies with extended PSFs were early detected by {\em Einstein}:
2E~1448 (NX~13) and 2E~1456 (NX~32, the second brightest X-ray
source in our sample). 
The NASA/IPAC Extragalactic Database (NED) tabulates isophotal major
axes $2a$ = 24.0 and 14.60\,arcsec at the reference level
20.0\,$K_{\rm s}$-mag\,arcsec$^{-2}$, respectively.
Both of them were catalogued as galaxies with peculiar colours by 
\citet{cab08c}:
on the one hand, NX~32 has very blue colours in the optical ($B_J - i \sim$
--0.3\,mag), but very red colours in the near infrared ($J-K_{\rm s}$ =
1.68$\pm$0.10\,mag). 
On the other hand, NX~13 has appreciable emission at 4.5--8.0\,$\mu$m
\citep{her07}. 
In Section.~\ref{section.xraygalaxies} we present spectral fitting of 
these two galaxies.

\subsubsection{Sources of unknown nature}
\label{section.unknown}

\paragraph{NX~11.}
We have not been able to discriminate which is the actual source of the X-ray
emission of NX~11.
As mentioned in Section~\ref{section.cross}, NX~11 has {\em two} possible
optical/near infrared counterparts.
One is a dwarf star candidate (up-triangle in right window in
Fig.~\ref{diagram.CCandCM}) and the other one (down-triangle) is a red galaxy
with an extended PSF.
Both counterparts are at $\sim 4$\,arcsec to the X-ray source.
If the dwarf star were the X-ray emitter, it would be an active star in
the field and not a $\sigma$~Orionis member candidate. 

\paragraph{NX~23.}
In spite of its clear detection in the PN, MOS1, and MOS2 cameras ($L$ = 79),
and its relative brightness in the optical and the near infrared (e.g. $J$ =
13.70$\pm$0.03\,mag), NX~23 displays the largest deviation between the
coordinates of the X-ray and optical/near infrared counterparts ($\rho \sim$
7\,arcsec). 
Besides, while the optical/near infrared counterpart have typical colours of
normal field dwarfs, the hardness ratios are peculiar for its hypothetical
stellar nature ($HR_1$ = --0.42$\pm$0.11, $HR_2$ = --0.62$\pm$0.11). 
To explain this dilemma, the origin of the X-ray source could be an X-ray
emitter galaxy in the background that is not detectable in the Digital Sky
Survey (nor DENIS nor 2MASS) instead of an irregular, extremely active, X-ray
emission from the correlated dwarf.

\paragraph{NX~41.}
This X-ray source was previously identified with a possible cluster candidate
from 2MASS by Franciosini et al. (2006; with identification number 14 in their
list).
However, it displays much bluer colours than confirmed $\sigma$~Orionis members
of the same magnitude ($J$ = 16.33$\pm$0.09\,mag, $i-K_{\rm s} \sim$ 1.8\,mag),
combined with soft hardness ratios ($HR_1$ = --0.47$\pm$0.16, $HR_2$ =
--0.48$\pm$0.16). It is probably another active galaxy in the background.

\subsection{Derivation of X-ray properties}
\label{section.properties}

\begin{table*}
\caption[]{X-ray properties of our 12 brightest X-ray sources associated
to $\sigma$~Orionis cluster members and candidates, with errors measured as the 
90\% confidence range.} 
\label{table.xray}
$$ 
\begin{tabular}{c l cc c c c c c}
\hline
\hline
\noalign{\smallskip}
NX 		& Mayrit	& $kT_1$ 			& $kT_2$ 			& $EM_1/EM_2$	& $Z/Z_\odot^{a}$ 		& $N_\mathrm{H}$ 		& $\chi^2_\mathrm{red}$/(d.o.f.) 	& $F_{\rm X}^{b}$ \\
  		&		& [keV] 			 & [keV]  			&		&             			& [$10^{21}$\,cm$^{-2}$]	& 					& [$10^{-13}$\,erg\,cm$^{-2}$\,s$^{-1}$] \\
\noalign{\smallskip}
\hline
\noalign{\smallskip}
5  		& 1415279AB	& 0.36$_{-0.04}^{+0.05}$	& 1.02$_{-0.06}^{+0.06}$ 	& 0.77		 & 0.13$_{-0.03}^{+0.04}$ 	& 0.56$_{-0.17}^{+0.19}$ 	& 0.98/155 				& 2.21$_{-2.12}^{+2.54}$ \\
\noalign{\smallskip}
7  		& 1374283	& 0.65$_{-0.16}^{+0.09}$ 	& 1.21$_{-0.73}^{+0.32}$ 	& 1.16 		& 0.18$_{-0.07}^{+0.11}$ 	& 1.04$_{-0.17}^{+0.37}$ 	& 1.27/98 				& 1.38$_{-1.14}^{+2.29}$ \\
\noalign{\smallskip}
14 		& 1149270	& 0.77$_{-0.15}^{+0.19}$ 	& ... 				& .... 		& 0.08$_{-0.03}^{+0.03}$ 	& 0.55$_{-0.31}^{+0.42}$ 	& 1.07/22 				& 0.09$_{-0.03}^{+0.13}$ \\
\noalign{\smallskip}
17 		& 1298302	& 0.90$_{-0.11}^{+0.15}$ 	& 3.31$_{-1.18}^{+60.7}$ 	& 0.36 		& 0.39$_{-0.14}^{+0.39}$ 	& 0.19$_{-0.19}^{+0.64}$ 	& 0.92/76 				& 1.28$_{-1.02}^{+1.46}$ \\
\noalign{\smallskip}
20 		& 1027277	& 0.75$_{-0.44}^{+0.18}$ 	& ...                      	& ...    	& 0.09$_{-0.09}^{+0.12}$	& 1.02$_{-0.81}^{+5.34}$ 	& 1.04/34 				& 0.12$_{-0.02}^{+0.02}$ \\
\noalign{\smallskip}
21 		& 1160240	& 0.72$_{-0.16}^{+0.14}$ 	& ...                      	& ...    	& 0.07$_{-0.02}^{+0.02}$ 	& 0.57$_{-0.27}^{+0.36}$ 	& 0.89/42 				& 0.23$_{-0.05}^{+0.04}$ \\
\noalign{\smallskip}
27$^{c}$	& 797272	& 0.32$_{-0.04}^{+0.06}$ 	& 1.12$_{-0.12}^{+0.15}$ 	& 0.84 		& 0.15$_{-0.08}^{+0.22}$ 	& 0.72$_{-0.22}^{+0.34}$ 	& 1.22/198 				& 3.47$_{-2.77}^{+3.74}$ \\
\noalign{\smallskip}
28 		& 789281	& 0.75$_{-0.06}^{+0.03}$ 	& 1.77$_{-0.21}^{+0.18}$ 	& 0.31 		& 0.38$_{-0.12}^{+0.12}$ 	& 0.46$_{-0.12}^{+0.16}$ 	& 1.10/336 				& 7.07$_{-6.78}^{+7.76}$ \\
\noalign{\smallskip}
30 		& 783254	& 0.78$_{-0.02}^{+0.02}$ 	& 1.92$_{-0.17}^{+0.13}$ 	& 0.60 		& 0.21$_{-0.06}^{+0.03}$ 	& 0.37$_{-0.06}^{+0.03}$ 	& 1.33/354 				& 11.6$_{-11.4}^{+12.4}$ \\
\noalign{\smallskip}
34 		& 887313	& 0.45$_{-0.33}^{+0.27}$ 	& ...                      	& ...    	& 0.03$_{-0.03}^{+0.10}$ 	& 1.18$_{-1.18}^{+5.67}$ 	& 1.13/27 				& 0.31$_{-0.01}^{+0.39}$ \\
\noalign{\smallskip}
37 		& 615296	& 0.34$_{-0.03}^{+0.04}$ 	& 1.14$_{-0.10}^{+0.07}$ 	& 0.63 		& 0.13$_{-0.03}^{+0.04}$ 	& 0.71$_{-0.15}^{+0.15}$ 	& 0.99/270 				& 4.49$_{-4.10}^{+4.96}$ \\
\noalign{\smallskip}
38 		& 547270	& 0.67$_{-0.19}^{+0.10}$ 	& 1.30$_{-0.30}^{+0.42}$ 	& 0.66 		& 0.51$_{-0.31}^{+0.84}$ 	& 0.17$_{-0.17}^{+0.07}$ 	& 1.04/48 				& 0.56$_{-0.39}^{+0.74}$ \\
\noalign{\smallskip}
\noalign{\smallskip}
\hline
\end{tabular}
$$ 
\begin{list}{}{}
\item[$^{a}$] Global abundance scaled on the solar value 
from \citet{and89}.
\item[$^{b}$] Unabsorbed X-ray flux in the 0.3--10.0\,keV band.
$F_{\rm X}$ was computed through fixing $N_\mathrm{H} = 0$ after 
fitting with XSPEC (using the APEC model). For details on the 
fitting procedure and how to calculate errors, we refer the reader to the 
online XSPEC guide (accessible from the HEASARC website: 
{\tt http://heasarc.nasa.gov/docs/xanadu/xspec/index.html}).
\item[$^{c}$] For source NX~27, a $3T$ model was also used. The third 
component has $kT_3 = 4.34_{-1.57}^{+59.2}$ keV and $EM_1/EM_3 = 1.01$.
\end{list}
\end{table*}

Of the 15 X-ray sources cross-identified with $\sigma$~Orionis members and
candidate members in Section~\ref{section.youngstars}, twelve have 
more than 800 counts in the summed EPIC image after background correction.
As a matter of a fact, all of them have maximum likelihood parameters $L
\gtrsim$ 100.
Only three young stars (NX~4, 16, and~36) have X-ray emission faint enough for
preventing the derivation of X-ray properties (i.e. less than about 400 counts,
$L \lesssim 25$).
We carried out a spectral analysis of the 12 brightest X-ray counterparts of
young stars using data from the EPIC camera and the  XSPEC spectral fitting
package \citep{arn96, arn04}.
The X-ray spectrum of each one of these sources was first corrected 
by subtracting a background spectrum, extracted from a close region in the 
same detector and scaled to the source extraction region area.
We adopted the Astrophysical Plasma Emission Database 
\citep[APED;][]{smi01}, 
that contains the relevant atomic data for  both continuum and
line emission included in the XSPEC software. 
Interstellar absorption, $N_\mathrm{H}$, was taken into account using the 
photo-electric absorption cross-sections of \citet{mor83}, 
also available in~XSPEC. 


In Table~\ref{table.xray}, we give the results of fitting with the 
Astrophysical Plasma Emission Code (APEC) model.
In general, we used a two-temperature ($2T$) model, with fixed 
relative abundances. 
We applied a one-temperature ($1T$) model only for the four sources 
with maximum likelihood parameters $L \sim$ 100 (i.e. less counts), since the
addition of a second thermal component did not improve the $\chi^{2}$
goodness-of-fit of the least-square solution.  
The remaining X-ray sources with satisfactory fitting to a $2T$ model have $L
\gtrsim$ 750 (and up to $L \sim$ 33\,000). 
For source NX~27, a third  thermal component was needed to obtain an
accurate fitting to the hard tail of its X-ray spectrum (a power-law component 
cannot satisfactorily fit it).
The temperatures and emission measure ratio given in the table correspond
to the soft and medium thermal components.
See further details in Section~\ref{section.variability}. 
In Figs.~\ref{nx05-21} and~\ref{nx27-38}, we show the X-ray spectral fitting of
the twelve young stars.

\section{Discussion}
\label{section.discussion}

\subsection{X-ray young stars}
\label{section.xrayyoungstars}

\subsubsection{New X-ray sources in $\sigma$~Orionis}
\label{section.newxrayyoungstars}


We have derived the X-ray properties of 12 young stars
(Table~\ref{table.xray}), while three of them were too faint.  
Of the latter stars,

\begin{itemize}

\item NX~4 is a bright Tycho-2 star with arguable X-ray emission
and expected spectral type at around late-A or early-F (with low frequencies of
X-ray emission), 
\item NX~16 is in the bluest part of the cluster sequence in the $i$ vs. $i-K_{\rm s}$
diagram, has a very low maximum likelihood parameter ($L$ = 17), and was only
detected by one camera (i.e. its youth is debatable), 
\item and NX~36, with reliable (faint) X-ray emission and $J-K_{\rm s}$ =
1.22$\pm$0.04\,mag, is the reddest young star in our sample and one of the
reddest stars in $\sigma$~Orionis. 
The infrared flux excess at IRAC and MIPS/{\em Spitzer} passbands and
its strong H$\alpha$ emission (see references in Table~\ref{table.class})
are an evidence of NX~36 (Mayrit~662301) having a developed disc, possibly
edge-on, that absorbs any possible X-ray emission from the corona or from
the channels that connect the inner border of the disc with the central object
(K\"onigl~1991). 
\end{itemize}

Of the twelve young stars with derived X-ray properties, four have not enough
signal --\,i.e. counts\,-- for an accurate fitting to a $2T$-model. 
Therefore, only one thermal component was fitted for the four of them.
They are: NX~20 (a confirmed young star with lithium), NX~14 and NX~21
(two new photometric cluster member candidates), and NX~34 (a non-variable
photometric cluster member candidate; Scholz \& Eisl\"offel 2004).
The X-ray emission ensures their extreme youth. 
Likewise, the set of eight young stars with at least two thermal components
comprises six confirmed stars with spectroscopic features of youth (NX~5, 7, 27,
28, 30, and~38) and two previously known photometric cluster member candidates
(NX~17 and~37).

To sum up, by deriving their X-ray properties, we show that five young star
candidates selected in colour-magnitude diagrams are actually young (two are
presented here for the first time --NX~14 and~21--; two were unconfirmed cluster
member candidates --NX~17 and~34--; one was confirmed by \citet{fra06}
--NX~37--). 
Besides, we first find X-ray emission from three young stars with lithium (NX~5,
7, and~20).  
In other words, of the 15 young stars investigated here, the detection of X-ray
emission of {\em nine} of them is new (but NX~4 and~16 detections are
questionable). 

Although these numbers are far from the numerous population of X-ray sources
first found in the trailblazer works by Wolk (1996) and \citet{fra06},
some of our young stars with emission are at previously uncovered
angular distances from the cluster centre.
Using previous {\em ROSAT} and {\em XMM-Newton} data, they could properly study
the innermost 15\,arcmin of $\sigma$~Orionis, and hardly in the 15--20\,arcmin
external region.
However, as shown by \citet{cab08a}, 
the ``core'' and the ``halo'' of the cluster extends up to 20 and 30\,arcmin, respectively.
At larger separations from the Trapezium-like system that defines the cluster
centre (i.e. $\rho >$ 30\,arcmin), the surface density of cluster members drops
notoriously, and the contamination by nearby young populations (around Mintaka to 
the northeast, Alnilam to the northwest, and the Horsehead Nebula to the southeast)
gets important \citep{jef06, cab08c}.
Except for NX~34 (that is at $\rho \sim$ 14.8\,arcmin from 
$\sigma$~Ori~AF--B),
all our X-ray young stars lie on the external (17\,arcmin $\lesssim 
\rho \lesssim$ 24\,arcmin) region (i.e. in the cluster core-to-halo 
transition).
The decrease in the surface density of X-ray young stars at large distances from
the cluster centre that we observe is an obvious consequence of the
decrease in the surface density of $\sigma$~Orionis members, that approximately
follows a distribution proportional to $\rho^{-1}$ 
\citep{cab08a}.

\subsubsection{Discs and $N_{\rm H}$}
\label{section.discs}

The value of the column density $N_{\rm H}$ in our X-ray model accounted
for both the interstellar column density and the circumstellar material that
could surround each star, like a protoplanetary disc or diffuse interstellar
material left over from its formation. 
On the one hand, from the low $E(B-V)$ colour excess towards the $\sigma$~Orionis
cluster, that lies in the interval 0.04--0.09\,mag \citep{lee68, bej01, bej04,
her05,she08,gon08},
it is derived that the
expected interstellar column density approximately varies between 2.2 and
4.9\,10$^{16}$\,m$^{-2}$ (i.e. $N_{\rm H} \approx$
0.22--0.49~10$^{21}$\,cm$^{-2}$). 
There are two of our 12 young stars with $N_{\rm H}$ larger than
this range ($N_{\rm H} \approx 1.0-1.2 \times 10^{21}$\,cm$^{-2}$), even 
taking the errors into account.
They are NX~7 ($J-K_{\rm s}$ = 0.48$\pm$0.03\,mag) and NX~34 ($J-K_{\rm s}$ =
0.96$\pm$0.05\,mag).
There are no hints for the first star to have a disc 
\citep[NX~7 falls out of the area investigated to date with the 
{\em Spitzer Space Telescope} and the Infrared Array Camera;][]{her07},
while the relatively red colour for the magnitude of the second star
might suggest the presence of a disc (NX~34 has the third reddest $J-K_{\rm s}$
among the 12~stars). 

On the other hand, there are two $\sigma$~Orionis stars in our analysis
catalogued as classical T~Tauri stars.
One is the faint X-ray emitter NX~36, that was discarded for the spectral energy
distribution fitting because of their low number of summed counts.
Its very red near-infrared colours ($J-K_{\rm s}$ = 1.22$\pm$0.04\,mag) 
support the scenario of an X-ray absorbent disc
(Section~\ref{section.newxrayyoungstars}).  
However, the other classical T~Tauri star with infrared flux excess at
2--8\,$\mu$m (NX~38, with colour index $J-K_{\rm s} = 1.10 \pm
0.04$\,mag) has a {\em low} value of $N_{\rm H}$.
There are four young stars whose X-ray parameters have been derived and possess
less counts than NX~38, and none of them displays such a $J-K_{\rm s}$-$N_{\rm
H}$ incompatibility.
Instead of claiming a contradiction with the scenario of an X-ray absorbent
disc, we propose a simpler solution based on a geometrical fact: 
the X-ray emission from the upper layers of the atmosphere of NX~38 passes
through a shorter portion of disc because of a relatively larger inclination
angle with our visual (i.e. the disc is pole-on).  
This hypothesis, nevertheless, should be verified by determining 
the disc inclination angle.  

\begin{figure}
\centering
\includegraphics[width=0.49\textwidth]{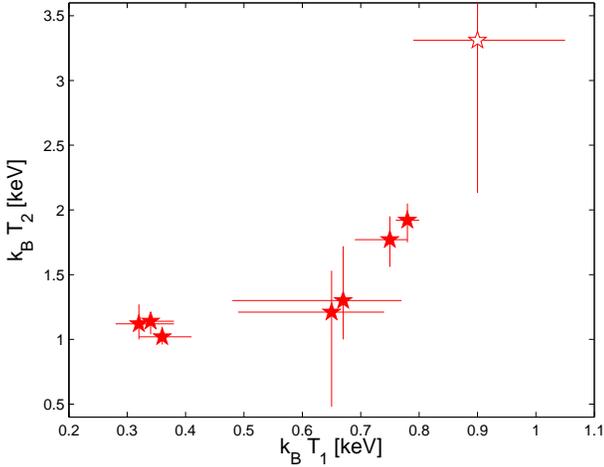}
\caption{Same as Fig.~\ref{diagram.CCandCM}, but for the $k_B T_2$ vs.
$k_B T_1$ diagram only for the young stars in Table~\ref{table.xray}.
The young star candidate with high temperatures is NX~17. Its upper limit 
of $k_B T_2$ is undetermined.}
\label{diagram.kBTs}
\end{figure}

\subsubsection{X-ray variability}
\label{section.variability}

\begin{figure*}
\centering
\includegraphics[width=8.3cm]{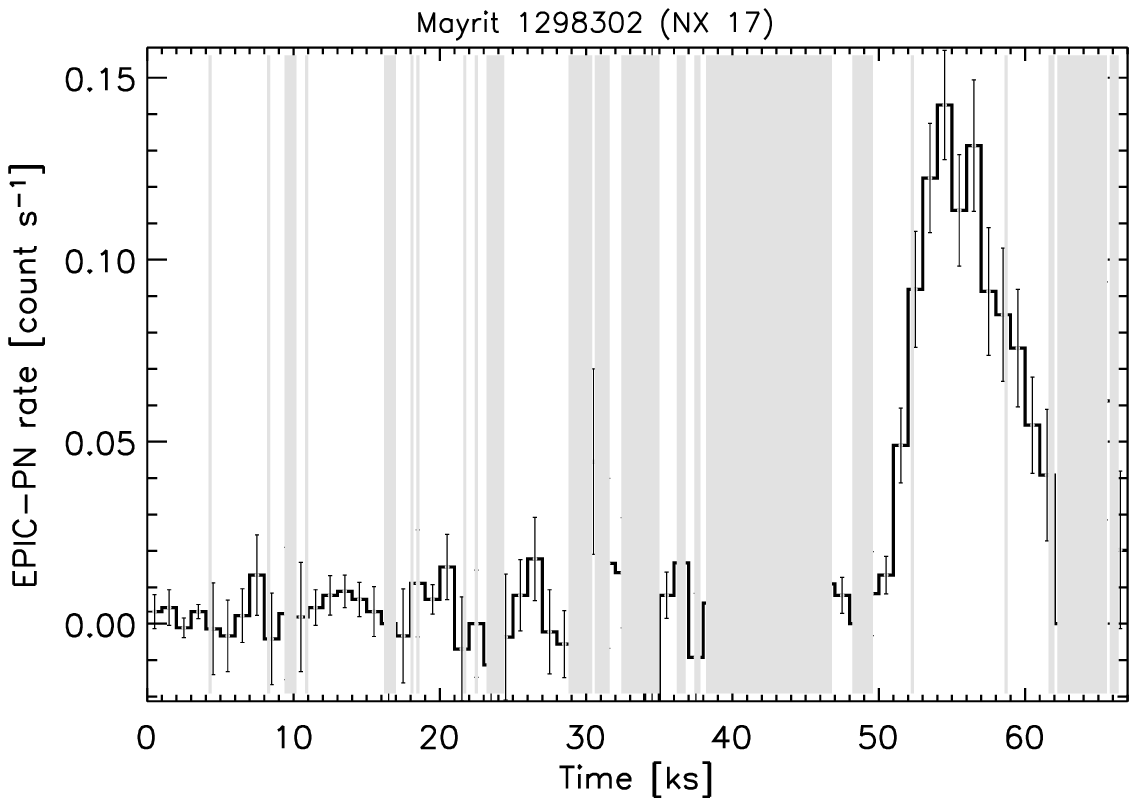}
\includegraphics[width=8.3cm]{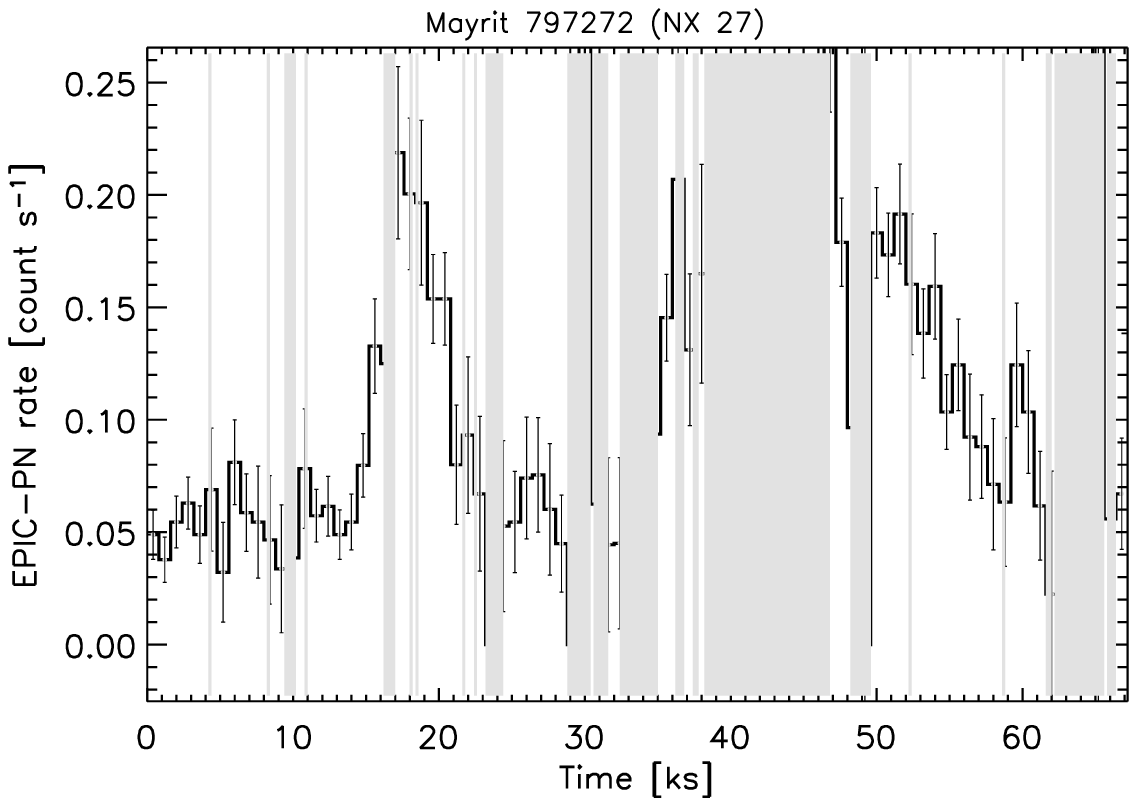}
\caption{Background subtracted X-ray lightcurves of the stars 
Mayrit 1298302 (NX\,17) and Mayrit 797272 (NX\,27) in the energy 
range 0.3--7.5\,keV. The bin size is 800\,s. The light grey area marks the 
parts of the spectrum rejected for this study because of the high background 
level. Both stars showed clear flaring events which
caused the rising of the coronal temperature (see text for details).}
\label{fig.curves}
\end{figure*}

In the source detection stage (Section~\ref{section.identification}), 
we created a {\em cleaned} event-list, rejecting 30\,\% of the observation 
(see Table~\ref{table.observations}). We used this event-list to look for 
variations in the lightcurve of our sources. Since the period in which 
high background level is present is large, the statistical analysis of 
variations was not possible. In these conditions, it is difficult to detect, 
for instance, variations due to rotational modulation. 
In our sample, 
two sources showed clear flaring events in their lightcurves: NX\,17 and 
NX\,27 (see Fig.~\ref{fig.curves}). In this section, we focus on such events. 

\paragraph{NX\,17.}
This source (identified with the young cluster member candidate 
Mayrit 1298302; see Table~\ref{table.class}) showed a large flare with 
a duration of more than $4$\,h and a relative increase in flux of $\sim 30$. 
This is much larger
than observed in young stars in Taurus \citep{fra07} and comparable 
with those observed in some young stellar objects in the Orion Nebula
Cluster \citep{fav05}.
The mean coronal temperature during the observation reached 3.3\,keV 
($T \sim 38$\,MK). This value is typical of high-energy flaring 
events in late-type stars \citep[see][and references therein]{cre07}.

\paragraph{NX\,27.}
This source has been identified with the $\sigma$~Orionis member Mayrit 797272,
an M1--3 type star that shows a large lithium absorption line 
\citep{wol96,cab08d}.
In its X-ray lightcurve, it showed a first flare at $t \approx 14$\,ks from 
the beginning of the observation and high variability during the second half of 
it, probably coming from a second flare. With a total duration of $\sim 9$\,ks,
the first flare presented a relative increase in flux of $\sim 4$, similar to 
the 
observed in other young stars \citep[e.g.][]{fav05,fra07} and in field M dwarfs
\citep[e.g.][]{rob05}. The temperature of the hotter component in the spectral 
fitting to the whole observation is 4.3\,keV ($T \sim 50$ MK). 
Both flares are probably contributing to the spectra. 

\subsection{Young stars and brown dwarfs without X-ray counterpart}

A series of star and bright brown dwarf members and candidates of 
the $\sigma$~Orionis cluster in the Mayrit catalogue \citep{cab08c}, and
that fall in the area covered by \textit{XMM-Newton}, 
were not detected in this observation because of their low X-ray emission. 
Most of them show features of youth, 
such as Li~{\sc i} in absorption, H$\alpha$ in strong broad emission, abnormal 
strength of alkali lines due to low gravity, early spectral types (OB), and/or flux 
excess longwards of the $J$ band (i.e. class~II). 
S\,Ori~55 (Section~\ref{section.introduction}), located in the centre of the
field of view and close to the central gaps, is one of these objects.
Other interesting ones are 
S\,Ori~70 (see, again, Section~\ref{section.introduction}), with a tentative 
mass of $\sim 3$\,$M_{\rm Jup}$ and a possible disc, and \object{S\,Ori~66}, 
also in the planetary-mass domain and with a (likely) mid-infrared flux excess
and strong H$\alpha$ emission \citep{bar01,zap08,sch08,luh08}. They are maybe 
the most interesting ultra low-mass objects in the area.
Some other candidates and members of the cluster fall in a detector gap 
or close to the border, which avoided any reliable detection of such sources.

For all the members and candidates of $\sigma$~Orionis in the Mayrit 
catalogue with no X-ray counterpart in our observation, we give  
upper limits to their X-ray emission in Table~\ref{table.upperlimits}.
Fluxes were determined from the count-rate measured in a circular region 
of 15\,arcsec centred on the position of each source in the (exposure-map 
corrected) EPIC image. The count-rate simulator 
WebPIMMS\footnote{WebPIMMS is available at:\\ 
{\tt http://heasarc.nasa.gov/Tools/w3pimms.html}.} was used to transform 
count-rates to fluxes.  We assumed a hot plasma model with $kT = 1$\,keV 
and a column density $N_\mathrm{H} = 0.27 \times 10^{21}$\,cm$^{-2}$ for 
every source (see Section~\ref{section.discs} for a detailed discussion on the 
interstellar column density in the direction of the $\sigma$~Orionis cluster).
The temperature of 1\,keV 
is representative of the corona of young stars and has been also observed
in the X-ray emitting brown dwarfs in the Orion Nebula cluster \citep{fei05}.
Some of the listed objects were marginally detected by the
detection algorithm, but showed maximum likelihood parameters below the value 
chosen here for reliable detection, $L \ge 15$.

The upper limits given here put constraints to the X-ray emission of the low 
mass objects in the field. 
Very few is known about the X-ray emission from brown dwarfs.
Apart from the $\sigma$~Orionis brown dwarfs introduced in
Section~\ref{section.introduction}, there have been only a handful of
detections in other young star-forming regions, such as the Orion Nebula
Cluster, Chamaeleon, and Taurus \citep{neu98,neu99,ima01,tsu03,pre05}.
There have also been reports of X-ray emission from a brown dwarf in the
Pleiades 
\citep{bri04} and two brown dwarfs in the field:
\object{LP~944--20} 
\citep[member of the Castor moving group,][]{rut00}
and \object{GJ~569B}[ab] 
\citep[questionable;][]{ste04}.
Nowadays, it is accepted that brown dwarfs and low-mass stars share the same
formation mechanism 
\citep[e.g.][]{cab07} and, therefore, brown dwarfs
are ``stars to scale'' without stable hydrogen burning.
Albeit brown dwarfs are smaller (fainter, cooler) than stars, they rotate much
faster 
\citep[with periods of about 10\,h; see][and references therein]{moh03} 
which may enhance their magnetically-driven, internal dynamos.
This enhancement might produce, in its turn, an extra heating of the hot,
tenuous, X-ray-emitting, upper atmospheric layer in brown dwarfs.

\subsection{Comparison to previous {\em XMM-Newton} observations}
\label{section.franciosini}

There are nine X-ray sources in common with the {\em XMM-Newton} observations
by \citet{fra06} and ours (see second column in Table~\ref{table.class}).
They are six young stars, two galaxies (NX~35 and~39), and a source of unknown
nature (NX~41; Section~\ref{section.unknown}), classified by \citet{fra06} 
as $\sigma$~Orionis cluster members or candidates, unidentified X-ray
sources, and a possible cluster candidate from 2MASS, respectively.
The X-ray parameters determined here for the stellar sources for which a 
spectral fitting was possible in both works (NX~28, NX~30, and NX~37) 
are well in agreement with those obtained by \citet{fra06}. The remaining 
stars in common are too faint in, at least, one of both works to perform a 
robust fitting to their spectra and, thus, we cannot compare results. The 
case of NX~27 
([FPS2006]\,1 in Franciosini et al. 2006) is remarkable since, in our 
observation, its count-rate is more than one order of magnitude higher than 
in the \citet{fra06}'s one. Here, the star showed high X-ray variability 
coming from large flaring events (Section~\ref{section.variability}) and
it had counts enough to do an accurate spectral fitting. The energy 
band used in our study, $0.3-7.5$\,keV, is similar to that 
of \citet{fra06}, who used $0.3-8.0$\,keV.
%

Besides, there are seven additional X-ray sources that \citet{fra06}
associated to
cluster members and candidates and that were not detected with our task,
although they fall in our survey area. However, all of them (with [FPS2006] 
identification number in the last column of Table~\ref{table.upperlimits}) are
very faint and located in (or very close to) borders or gaps of 
the detectors in our observation.

\subsection{X-ray galaxies}
\label{section.xraygalaxies}

Of the 12 X-ray sources in Table~\ref{table.sources} with maximum likelihood
parameters $L \gtrsim 400$, eight are young stars and four are background
galaxies with $HR_2 > -0.5$. 
The remaining 29 sources have $L \lesssim 100$. Two of the four hard X-ray 
emitters in our sample (NX~13 and~32) were previously detected with 
{\em Einstein} and identified with galaxies (see Section~\ref{section.galaxies}
for details). They are the eighth and second brightest sources in this 
{\em XMM-Newton} field, respectively. The other two galaxies are NX~12 and~26 
(with $L$ = 1601 and 
394, respectively) and have no near-infrared counterpart (NX~26 has no optical
counterpart, either).
 
The study of the X-ray properties of the AGNs in the field is far from the 
scope of this paper. Nevertheless, we show in Fig.~\ref{fig.agns} the X-ray 
spectra observed here for NX\,13 and NX\,32, together with a simple fitting 
using a power-law, to compare with the spectra of the young stars 
(Figs.\,\ref{nx05-21} and \ref{nx27-38}). The X-ray spectra of these sources 
are clearly distinguishable from those of the young stars in our sample.


\section{Summary and final remarks}
\label{section.summary}

We used archived data from the {\em XMM-Newton} mission to explore a region 
to the west of the centre of the $\sigma$~Orionis cluster, slightly overlapped
with a previous observation centred on the eponymous stellar system $\sigma$ 
Ori. Our aim was to search for new X-ray young stars in the region and 
determine their X-ray properties. After a careful inspection of the source list
resultant from the use of a maximum likelihood technique to reveal sources in
the images of the EPIC cameras, we kept 41 X-ray sources. Among them, a total 
of 15 are young stars, 4 are field stars, and 19 are AGNs. The remaining three
sources are of unknown nature, probably AGNs given their hardness ratios.
Many of the revealed X-ray sources, including those associated to young stars,
are new.

We derived X-ray parameters of the twelve young stars with enough counts
for a spectral fitting, using multi-temperature models with absorption. In
particular, we determined the column density, temperature of the thermal
components, and metallicities.
%
%
We observed no relation among the stars showing infrared excess and the
column density measured here.
Although we could not perform a 
detailed study of the X-ray variability because of the high variable background 
affecting the observation, two clear flaring events in the 
M-type young stars NX~17 (Mayrit 1298302) and NX~27 (Mayrit 797272) 
were studied. In both cases, we determined high coronal temperatures, 
reaching $38$ and $50$\,MK, respectively. 
We also determined upper limits for the X-ray flux of young 
stars and brown dwarfs in the field that were not detected by the searching 
algorithm used by us. 

Deep X-ray observations, combined with photometric data, have historically 
revealed as a powerful tool for disclosing the stellar population 
of star-forming regions down to the substellar boundary. In particular, in 
$\sigma$~Orionis we were able to detect X-ray emission of mid-M type stars.
Our work complements previous investigations on the X-ray population 
of this cluster. 
Our study also advances in the knowledge of the X-ray properties of 
young stars in the external region of the cluster, where the stellar density
drops. Deeper X-ray observations in this region should permit to 
detect also the X-ray counterparts of brown dwarfs 
(as it was done in the cluster centre). The combination of their X-ray 
properties with the results of near- and mid-infrared photometry (to reveal the 
presence of discs) will permit to unveil the origin of the X-ray emission of 
these objects (whether it is produced by accretion or by a hot corona, as 
in pre-main sequence stars). This work should be followed by the study of 
the frequency of brown dwarfs with and without disc showing X-ray emission.
A detailed study on the X-ray luminosity function of 
$\sigma$~Orionis, compared with other young stellar clusters, will 
be addressed in a forthcoming paper by the authors.


\begin{acknowledgements}

We thank the anonymous referee for helpful comments.
J.L.S. is an AstroCAM post-doctoral fellow at the~UCM. 
J.A.C. is an Investigador Juan de la Cierva at the~UCM. 
Partial financial support was provided by the Universidad Complutense de Madrid
and the Spanish Ministerio Educaci\'on y Ciencia and the European 
Social Fund under grant 
AYA2005--02750 of the Programa Nacional de Astronom\'{\i}a y Astrof\'{\i}sica
and by the Comunidad Aut\'onoma de Madrid under PRICIT project 
S--0505/ESP--0237 (AstroCAM). J.L.S. also acknowledges financial contribution 
by the Marie Curie Actions grant No. MERG-CT-2007-046535.

\end{acknowledgements}

\appendix

\section{On-line material}

\begin{table*}
\caption[]{Detected X-ray sources in the combined images of
the EPIC cameras.}
\label{table.sources}
\begin{tabular}{c cc c ccc cc}
\hline
\hline
\noalign{\smallskip}
NX    & \multicolumn{1}{c}{$\alpha$} & \multicolumn{1}{c}{$\delta$} &
        \multicolumn{1}{c}{$L$} & \multicolumn{3}{c}{$CR$ [ks$^{-1}$]} &
        \multicolumn{1}{c}{$HR_1$} & \multicolumn{1}{c}{$HR_2$} \\
      \noalign{\smallskip}
      \cline{5-7}
      \noalign{\smallskip}
      & \multicolumn{1}{c}{(J2000)} & \multicolumn{1}{c}{(J2000)} &   &
        \multicolumn{1}{c}{PN} & \multicolumn{1}{c}{MOS1} &
        \multicolumn{1}{c}{MOS2} &         &        \\
\noalign{\smallskip}
\hline
\noalign{\smallskip}
  1 &  05 36 46.30 & --02 35 04.0 &    27 &   9.8 $\pm$ 1.8 & ...	      &   3.8 $\pm$ 0.7 & --0.76 $\pm$ 0.20 &  +0.15 $\pm$ 0.20 \\
  2 &  05 36 50.88 & --02 24 52.5 &    45 &  15.3 $\pm$ 2.2 &	4.8 $\pm$ 1.0 &   6.1 $\pm$ 0.9 & --0.40 $\pm$ 0.13 & --1.00 $\pm$ 0.13 \\
  3 &  05 36 56.28 & --02 42 12.2 &    16 & ... 	    & ...	      &   4.3 $\pm$ 0.9 &  +0.08 $\pm$ 0.21 & --0.45 $\pm$ 0.21 \\
  4 &  05 37 10.25 & --02 30 09.5 &    15 &   4.3 $\pm$ 1.0 & ...	      & ...		& --0.81 $\pm$ 0.21 &  +0.14 $\pm$ 0.21 \\
  5 &  05 37 11.46 & --02 32 10.6 &  5375 & 114.0 $\pm$ 2.3 &  30.7 $\pm$ 1.0 &  31.3 $\pm$ 1.0 & --0.94 $\pm$ 0.01 & --1.00 $\pm$ 0.01 \\
  6 &  05 37 13.81 & --02 43 54.2 &    18 &   8.6 $\pm$ 1.9 & ...	      &   3.2 $\pm$ 0.7 & --0.68 $\pm$ 0.22 & --0.26 $\pm$ 0.22 \\
  7 &  05 37 15.20 & --02 30 55.5 &  2235 &  61.1 $\pm$ 1.8 &  17.0 $\pm$ 0.8 &  19.6 $\pm$ 0.8 & --0.89 $\pm$ 0.02 & --0.90 $\pm$ 0.02 \\
  8 &  05 37 19.63 & --02 25 41.7 &    24 &   6.7 $\pm$ 1.2 & ...	      & ...		& --0.51 $\pm$ 0.17 & --1.00 $\pm$ 0.17 \\
  9 &  05 37 20.91 & --02 37 20.4 &    21 &   3.7 $\pm$ 0.7 & ...	      &   1.8 $\pm$ 0.4 & --1.00 $\pm$ 0.07 & ...		\\
 10 &  05 37 22.75 & --02 22 42.5 &    19 &   7.7 $\pm$ 1.5 & ...	      &   3.2 $\pm$ 0.6 & --0.34 $\pm$ 0.18 & --0.49 $\pm$ 0.18 \\
 11 &  05 37 22.81 & --02 32 47.2 &    27 & ... 	    &	1.7 $\pm$ 0.3 &   1.3 $\pm$ 0.3 & --0.81 $\pm$ 0.19 & --0.05 $\pm$ 0.19 \\
 12 &  05 37 23.95 & --02 42 01.1 &  1601 &  88.1 $\pm$ 2.8 &  32.9 $\pm$ 1.3 &  32.8 $\pm$ 1.6 & --0.54 $\pm$ 0.03 & --0.26 $\pm$ 0.03 \\
 13 &  05 37 27.84 & --02 40 04.0 &  2597 &  ... 	    &  57.0 $\pm$ 1.8 &  49.5 $\pm$ 1.6 & --0.59 $\pm$ 0.03 & --0.70 $\pm$ 0.03 \\
 14 &  05 37 28.00 & --02 36 09.1 &    91 &   7.1 $\pm$ 0.8 &	1.7 $\pm$ 0.3 & ...		& --0.99 $\pm$ 0.06 & --1.00 $\pm$ 0.06 \\
 15 &  05 37 28.29 & --02 32 45.8 &    63 &   6.4 $\pm$ 0.8 &	1.7 $\pm$ 0.3 &   2.2 $\pm$ 0.3 & --0.67 $\pm$ 0.12 & --0.51 $\pm$ 0.12 \\
 16 &  05 37 28.36 & --02 24 19.9 &    17 & ... 	    &	1.8 $\pm$ 0.4 & ...		& --0.92 $\pm$ 0.18 & --1.00 $\pm$ 0.18 \\
 17 &  05 37 31.30 & --02 24 28.9 &  1422 &  61.2 $\pm$ 2.2 &  15.7 $\pm$ 1.0 &  16.1 $\pm$ 0.9 & --0.81 $\pm$ 0.03 & --0.60 $\pm$ 0.03 \\
 18 &  05 37 31.53 & --02 25 12.0 &    58 &   8.7 $\pm$ 1.2 &	2.1 $\pm$ 0.5 &   1.9 $\pm$ 0.4 & --0.62 $\pm$ 0.13 & --0.58 $\pm$ 0.13 \\
 19 &  05 37 32.95 & --02 37 45.6 &    35 &   5.8 $\pm$ 0.9 &	2.0 $\pm$ 0.4 &   2.3 $\pm$ 0.5 & --0.47 $\pm$ 0.15 & --0.70 $\pm$ 0.15 \\
 20 &  05 37 36.47 & --02 34 02.7 &   122 &  10.0 $\pm$ 1.0 &	2.8 $\pm$ 0.4 &   3.3 $\pm$ 0.4 & --0.87 $\pm$ 0.09 & --1.00 $\pm$ 0.09 \\
 21 &  05 37 37.67 & --02 45 45.1 &   102 &  18.1 $\pm$ 2.0 &	5.2 $\pm$ 0.8 &   6.2 $\pm$ 0.9 & --0.84 $\pm$ 0.10 & --1.00 $\pm$ 0.10 \\
 22 &  05 37 41.58 & --02 29 10.8 &    43 &   5.0 $\pm$ 0.8 & ...	      &   1.3 $\pm$ 0.3 & --0.88 $\pm$ 0.14 & --1.00 $\pm$ 0.14 \\
 23 &  05 37 41.84 & --02 29 39.8 &    79 &   9.2 $\pm$ 1.0 &	2.9 $\pm$ 0.4 &   2.7 $\pm$ 0.4 & --0.42 $\pm$ 0.11 & --0.62 $\pm$ 0.11 \\
 24 &  05 37 42.71 & --02 25 13.5 &    23 &  19.8 $\pm$ 3.8 &	2.5 $\pm$ 0.5 &   3.1 $\pm$ 0.5 & --0.79 $\pm$ 0.21 &  +0.37 $\pm$ 0.21 \\
 25 &  05 37 43.38 & --02 27 53.6 &    21 & ... 	    & ...	      &   2.2 $\pm$ 0.4 &  +0.07 $\pm$ 0.19 & --0.40 $\pm$ 0.19 \\
 26 &  05 37 47.56 & --02 29 10.2 &   394 &  24.4 $\pm$ 1.3 &	9.1 $\pm$ 0.6 &   9.3 $\pm$ 0.6 & --0.27 $\pm$ 0.05 & --0.45 $\pm$ 0.05 \\
 27 &  05 37 51.45 & --02 35 26.2 &  6907 & 136.8 $\pm$ 2.5 &  39.2 $\pm$ 1.4 &  40.6 $\pm$ 1.1 & --0.82 $\pm$ 0.01 & --0.58 $\pm$ 0.01 \\
 28 &  05 37 52.83 & --02 33 34.9 &  2643 & 322.5 $\pm$ 8.2 &  98.0 $\pm$ 1.7 & ...		& --0.85 $\pm$ 0.02 & --0.61 $\pm$ 0.02 \\
 29 &  05 37 53.27 & --02 26 54.7 &    21 &  13.0 $\pm$ 2.5 &	2.6 $\pm$ 0.6 & ...		& --0.70 $\pm$ 0.20 &  +0.26 $\pm$ 0.20 \\
 30 &  05 37 54.25 & --02 39 30.8 & 32606 & 547.1 $\pm$ 5.3 & 153.8 $\pm$ 2.4 & 157.6 $\pm$ 2.5 & --0.85 $\pm$ 0.01 & --0.70 $\pm$ 0.01 \\
 31 &  05 37 55.74 & --02 33 43.0 &    17 & ... 	    &	4.1 $\pm$ 0.8 & ...		& --0.90 $\pm$ 0.16 & --1.00 $\pm$ 0.16 \\
 32 &  05 37 56.20 & --02 45 13.8 & 26045 & 883.0 $\pm$ 8.9 & ...	      & 282.3 $\pm$ 4.3 & --0.60 $\pm$ 0.01 & --0.40 $\pm$ 0.01 \\
 33 &  05 37 58.07 & --02 25 14.0 &    34 &   8.8 $\pm$ 1.4 &	3.2 $\pm$ 0.6 &   4.0 $\pm$ 0.6 & --0.42 $\pm$ 0.15 & --0.42 $\pm$ 0.15 \\
 34 &  05 38 01.43 & --02 25 54.2 &    95 &  12.1 $\pm$ 1.4 &	4.6 $\pm$ 0.6 &   4.1 $\pm$ 0.6 & --0.84 $\pm$ 0.10 & --1.00 $\pm$ 0.10 \\
 35 &  05 38 03.68 & --02 27 30.0 &    16 &   4.8 $\pm$ 1.0 & ...	      & ...		& --0.65 $\pm$ 0.22 & --1.00 $\pm$ 0.22 \\
 36 &  05 38 06.57 & --02 30 23.9 &    24 &   5.3 $\pm$ 1.0 &	2.3 $\pm$ 0.5 & ...		& --0.82 $\pm$ 0.18 & --0.85 $\pm$ 0.18 \\
 37 &  05 38 07.65 & --02 31 32.3 &  8021 & 205.0 $\pm$ 3.7 &  62.9 $\pm$ 1.9 &  57.9 $\pm$ 1.5 & --0.89 $\pm$ 0.01 & --0.80 $\pm$ 0.01 \\
 38 &  05 38 08.06 & --02 35 57.3 &   749 &  38.7 $\pm$ 1.8 &  11.4 $\pm$ 0.8 &  10.6 $\pm$ 0.8 & --0.85 $\pm$ 0.04 & --0.52 $\pm$ 0.04 \\
 39 &  05 38 13.60 & --02 35 09.7 &   143 &  16.3 $\pm$ 1.5 &	5.7 $\pm$ 0.7 &   4.8 $\pm$ 0.7 & --0.67 $\pm$ 0.08 & --0.81 $\pm$ 0.08 \\
 40 &  05 38 14.02 & --02 36 48.3 &    15 & ... 	    &	2.8 $\pm$ 0.6 &   4.6 $\pm$ 0.7 & --0.06 $\pm$ 0.22 & --0.34 $\pm$ 0.22 \\
 41 &  05 38 15.42 & --02 42 11.7 &    30 &  12.4 $\pm$ 2.1 &	5.4 $\pm$ 1.0 & ...		& --0.47 $\pm$ 0.16 & --0.48 $\pm$ 0.16 \\
\noalign{\smallskip}
\hline
\end{tabular}
\end{table*}

\begin{table*}
\caption[]{USNO-B1/DENIS/2MASS counterparts of the X-ray sources in
Table~\ref{table.sources}.}
\label{table.counterparts}
\begin{tabular}{c cc ccccc l}
\hline
\hline
\noalign{\smallskip}
NX    	& $\alpha$	& $\delta$ 	& $B_J$		& $i$		& $J \pm \delta J$	& $H \pm \delta H$	& $K_{\rm s} \pm \delta K_{\rm s}$	& Alternative 	\\
    	& (J2000)	& (J2000) 	& [mag]		& [mag]		& [mag]			& [mag]			& [mag]					& name		\\
\noalign{\smallskip}
\hline
\noalign{\smallskip}
  1 	&  ...	 	&  ...	 	& $\gtrsim$21	& $\gtrsim$18.0	& $\gtrsim$17.1		& $\gtrsim$16.4		& $\gtrsim$14.3				& ...					\\ 
  2 	&  ...	 	&  ...	 	& $\gtrsim$21	& $\gtrsim$18.0	& $\gtrsim$17.1		& $\gtrsim$16.4		& $\gtrsim$14.3				& ...					\\ 
  3 	&  05 36 56.29 	& --02 42 07.4 	& 19.9		& 18.6		& $\gtrsim$17.1		& $\gtrsim$16.4		& $\gtrsim$14.3				& USNO--B1.0 0872--0105742 		\\ 
  4 	&  05 37 10.47 	& --02 30 07.2 	& 12.23$\pm$0.16& 11.31		& 10.72$\pm$0.05	& 10.46$\pm$0.05	&  10.30$\pm$0.04			& \object{TYC 4770--1261--1} 		\\ 
  5 	&  05 37 11.61 	& --02 32 08.8 	& 13.6		& 11.65		& 10.88$\pm$0.02	& 10.29$\pm$0.02	&  10.07$\pm$0.02			& \object{OriNTT 429} AB		\\ 
  6 	&  05 37 13.92 	& --02 43 51.7 	& 16.2		& 14.32		& 13.50$\pm$0.03	& 12.81$\pm$0.03	&  12.74$\pm$0.03			& 2MASS J05371392--0243517 	 	\\ 
  7 	&  05 37 15.37 	& --02 30 53.4 	& 13.5		& 12.18		& 11.36$\pm$0.02	& 11.01$\pm$0.02	&  10.88$\pm$0.02			& \object{SO211394}			\\ 
  8 	&  ...	 	&  ...	 	& $\gtrsim$21	& $\gtrsim$18.0	& $\gtrsim$17.1		& $\gtrsim$16.4		& $\gtrsim$14.3				& ...					\\ 
  9 	&  05 37 21.07 	& --02 37 19.2 	& 15.9		& 14.25		& 13.27$\pm$0.03	& 12.76$\pm$0.03	&  12.60$\pm$0.03			& 2MASS J05372107--0237192 		\\ 
 10 	&  05 37 23.05 	& --02 22 43.1 	& 16.9		& 16.28		& 15.46$\pm$0.06	& 15.07$\pm$0.10	&  14.88$\pm$0.14			& 2MASS J05372304--0222431 		\\ 
 11 	&  05 37 22.78 	& --02 32 42.9 	& 16.6		& 15.26		& 14.36$\pm$0.04	& 13.84$\pm$0.04	&  13.82$\pm$0.06			& 2MASS J05372277--0232429 		\\ 
  	&  05 37 23.06 	& --02 32 46.6 	& 19.0		& 15.78		& 14.18$\pm$0.04	& 13.59$\pm$0.04	&  13.41$\pm$0.05			& 2MASS J05372306--0232465 	 	\\ 
 12 	&  05 37 24.11 	& --02 41 59.9 	& 18.9		& 18.41		& $\gtrsim$17.1		& $\gtrsim$16.4		& $\gtrsim$14.3				& DENIS J05372411--0241599 		\\ 
 13 	&  05 37 27.94 	& --02 40 03.2 	& 15.8		& 16.19		& 14.82$\pm$0.08	& 14.09$\pm$0.07	&  14.43$\pm$0.07			& \object{2E 1448}			\\ 
 14	&  05 37 28.07  & --02 36 06.6  & 18.5  	& 15.27 	& 13.74$\pm$0.03	& 13.08$\pm$0.03	&  12.80$\pm$0.03			& 2MASS J05372806--0236065		\\ 
 15	&  ...  	&  ...  	& $\gtrsim$21	& $\gtrsim$18.0 & $\gtrsim$17.1 	& $\gtrsim$16.4 	& $\gtrsim$14.3 			& ...					\\ 
 16	&  05 37 28.32  & --02 24 18.2  & 18.6  	& 15.61 	& 14.00$\pm$0.03	& 13.39$\pm$0.03	&  13.08$\pm$0.03			& 2MASS J05372831--0224182		\\ 
 17	&  05 37 31.54  & --02 24 27.0  & 16.5  	& 13.63 	& 12.11$\pm$0.03	& 11.36$\pm$0.02	&  11.17$\pm$0.02			& \object{[SWW2004] 137}		\\ 
 18	&  05 37 31.68  & --02 25 10.6  & 19.7  	& 18.8  	& $\gtrsim$17.1 	& $\gtrsim$16.4 	& $\gtrsim$14.3 			& USNO--B1.0 0875--0102588		\\ 
 19	&  05 37 32.97  & --02 37 44.2  & 20.7  	& $\gtrsim$18.0 & $\gtrsim$17.1 	& $\gtrsim$16.4 	& $\gtrsim$14.3 			& USNO--B1.0 0873--0105537		\\ 
 20	&  05 37 36.67  & --02 34 00.3  & 17.5  	& 14.56 	& 13.00$\pm$0.03	& 12.30$\pm$0.02	&  12.05$\pm$0.02			& \object{[SWW2004] 141}		\\ 
 21	&  05 37 37.84  & --02 45 44.2  & 17.5  	& 14.17 	& 12.69$\pm$0.03	& 11.94$\pm$0.02	&  11.72$\pm$0.03			& 2MASS J05373784--0245442		\\ 
 22	&  05 37 41.79  & --02 29 08.1  & 11.23$\pm$0.06& 10.39 	&  9.80$\pm$0.03	&  9.60$\pm$0.03	&   9.50$\pm$0.03			& \object{TYC 4771--621--1}		\\ 
 23	&  05 37 41.61  & --02 29 38.0  & 16.5  	& 14.85 	& 13.70$\pm$0.03	& 13.11$\pm$0.02	&  12.94$\pm$0.03			& 2MASS J05374160--0229380		\\ 
 24	&  05 37 43.10  & --02 25 13.1  & 15.1  	& 13.46 	& 12.54$\pm$0.03	& 12.03$\pm$0.02	&  11.97$\pm$0.03			& 2MASS J05374310--0225131		\\ 
 25	&  ...  	&  ...  	& $\gtrsim$21	& $\gtrsim$18.0 & $\gtrsim$17.1 	& $\gtrsim$16.4 	& $\gtrsim$14.3 			& ...					\\ 
 26	&  ...  	&  ...  	& $\gtrsim$21	& $\gtrsim$18.0 & $\gtrsim$17.1 	& $\gtrsim$16.4 	& $\gtrsim$14.3 			& ...					\\ 
 27	&  05 37 51.61  & --02 35 25.7  & 16.6  	& 13.45 	& 11.89$\pm$0.03	& 11.17$\pm$0.02	&  10.98$\pm$0.02			& \object{[SWW2004] 125}		\\ 
 28	&  05 37 53.03  & --02 33 34.4  & 12.4  	& 10.80 	&  9.99$\pm$0.03	&  9.60$\pm$0.02	&   9.47$\pm$0.02			& \object{2E 1454}			\\ 
 29	&  ...  	&  ...  	& $\gtrsim$21	& $\gtrsim$18.0 & $\gtrsim$17.1 	& $\gtrsim$16.4 	& $\gtrsim$14.3 			& ...					\\ 
 30	&  05 37 54.40  & --02 39 29.8  & 12.02$\pm$0.12& 10.27 	&  9.26$\pm$0.02	&  8.72$\pm$0.05	&   8.61$\pm$0.02			& \object{2E 1455}			\\ 
 31	&  ...  	&  ...  	& $\gtrsim$21	& $\gtrsim$18.0 & $\gtrsim$17.1 	& $\gtrsim$16.4 	& $\gtrsim$14.3 			& ...					\\ 
 32	&  05 37 56.31  & --02 45 13.1  & 16.4  	& 16.73 	& 15.40$\pm$0.07	& 14.52$\pm$0.08	&  13.72$\pm$0.07			& \object{2E 1456}			\\ 
 33	&  05 37 58.36  & --02 25 13.1  & 19.7  	& $\gtrsim$18.0 & $\gtrsim$17.1 	& $\gtrsim$16.4 	& $\gtrsim$14.3 			& USNO--B1.0 0875--0102882		\\ 
 34	&  05 38 01.67  & --02 25 52.7  & 17.5  	& 14.65 	& 13.03$\pm$0.03	& 12.32$\pm$0.02	&  12.07$\pm$0.03			& \object{[SE2004] 53}  		\\ 
 35	&  ...  	&  ...  	& $\gtrsim$21	& $\gtrsim$18.0 & $\gtrsim$17.1 	& $\gtrsim$16.4 	& $\gtrsim$14.3 			& ...					\\ 
 36	&  05 38 06.74  & --02 30 22.8  & 15.8  	& 13.23 	& 11.76$\pm$0.03	& 10.92$\pm$0.02	&  10.54$\pm$0.02			& \object{Kiso A--0904 67}		\\ 
 37	&  05 38 07.85  & --02 31 31.4  & 13.3  	& 11.63 	& 10.57$\pm$0.03	&  9.93$\pm$0.02	&   9.77$\pm$0.02			& \object{2E 1459}			\\ 
 38	&  05 38 08.27  & --02 35 56.3  & 16.3  	& 13.97 	& 12.14$\pm$0.03	& 11.38$\pm$0.02	& 11.047$\pm$0.019			& \object{Kiso A--0976 316}		\\ 
 39	&  ...  	&  ...  	& $\gtrsim$21	& $\gtrsim$18.0 & $\gtrsim$17.1 	& $\gtrsim$16.4 	& $\gtrsim$14.3 			& ...					\\ 
 40	&  ...  	&  ...  	& $\gtrsim$21	& $\gtrsim$18.0 & $\gtrsim$17.1 	& $\gtrsim$16.4 	& $\gtrsim$14.3 			& ...					\\ 
 41 	&  05 38 15.53 	& --02 42 05.1 	& 18.0		& 17.25		& 16.33$\pm$0.09	& 15.61$\pm$0.09	& 15.5:					& \object{2MASS J05381552--0242051}	\\ 
\noalign{\smallskip}
\hline
\end{tabular}
\end{table*}

\begin{table*}
\caption[]{Upper limits ($3\sigma$) for undetected 
           $\sigma$~Orionis cluster members and candidates in the
           {\em XMM-Newton} field.} 
\label{table.upperlimits}
$$ 
\scriptsize
\begin{tabular}{llcclcl}
\hline
\hline
\noalign{\smallskip}
Mayrit 	& Alternative			& $\alpha$	& $\delta$	& Youth					& F$_{X}^{a}$ & Remarks\\ 
	& name				& (J2000)	& (J2000)	& features				& [10$^{-13}$ erg cm$^{-2}$ s$^{-1}$]  &   \\ 
\noalign{\smallskip}
\hline
\noalign{\smallskip}
1773275	& [SWW2004] 13			& 05 36 46.91   & --02 33 28.3  & Li~{\sc i}, low~$g$			& ...     & In PN gap					\\ 
1599271	& S\,Ori 33			& 05 36 58.08	& --02 35 19.4	& ...					& $<$0.29 & ... 						\\ 
1630286	& [SWW2004] 90			& 05 37 00.30	& --02 28 26.6	& ...					& $<$0.33 & ... 						\\ 
1738301	& ...				& 05 37 05.17	& --02 21 09.4	& ...					& ...     & In MOS border. Out of PN field		\\ 
1416280	& [SWW2004] 22			& 05 37 11.68	& --02 31 56.7	& ...					& ...     & Blended with NX 5 				\\ 
1391255	& S\,Ori J053715.1--024202	& 05 37 15.16	& --02 42 01.6	& Li~{\sc i}, H$\alpha$			& $<$0.27 & ... 						\\ 
1225282 & S\,Ori 66			& 05 37 24.70   & --02 31 52.0  & H$\alpha$, II?			& $<$0.26 & ... \\
1185274 & S\,Ori 55			& 05 37 25.90   & --02 34 32.0  & H$\alpha$, II?                        & $<$0.22 & ... \\
1329304 & Haro 5--5			& 05 37 30.95   & --02 23 42.8  & H$\alpha$, II				& $<$0.29 & Marginally detected				\\ 
1227243	& HD 294275			& 05 37 31.87	& --02 45 18.5 	& OB					& ...     & In PN gap					\\ 
1176297 & [HHM2007] 107  		& 05 37 35.14   & --02 26 57.7  & ...					& $<$0.23 & ... 						\\ 
1116300	& HD 37333			& 05 37 40.48	& --02 26 36.8 	& OB, Si~{\sc ii}			& $<$0.19 & ... 						\\ 
968292	& GSC 04771--00962		& 05 37 44.92	& --02 29 57.3 	& ...					& $<$0.22 & ... 						\\ 
958292	& SO210868			& 05 37 45.57	& --02 29 58.5 	& Li~{\sc i}				& ...     & In PN gap					\\ 
999306  & S\,Ori 23			& 05 37 51.11   & --02 26 07.5  & ...					& $<$0.25 & ... 						\\ 
790270	& [KJN2005] 62			& 05 37 52.07	& --02 36 04.7	& Li~{\sc i}, low~$g$			& $<$0.26 & Marginally detected				\\ 
882239	& ...				& 05 37 54.45	& --02 43 37.8	& ...					& $<$0.33 & ... 						\\ 
809248	& [SWW2004] 174			& 05 37 54.86	& --02 41 09.2	& Li~{\sc i}, low~$g$			& $<$0.37 & Marginally detected				\\ 
757283	& S\,Ori 35			& 05 37 55.60	& --02 33 05.3	& II					& $<$0.23 & ... 						\\ 
728257	& S\,Ori 12			& 05 37 57.46	& --02 38 44.4	& Li~{\sc i}, H$\alpha$, low~$g$, II	& $<$0.26 & ... 						\\ 
767245	& [OJV2004] 28			& 05 37 58.40	& --02 41 26.2	& Li~{\sc i}, low~$g$			& $<$0.28 & ... 						\\ 
861230	& [SWW2004] 140			& 05 38 00.56   & --02 45 09.7  & ...					& ...     & [FPS2006] 4. Close to PN border and to NX 32 \\ 
884312	& [WB2004] 13			& 05 38 00.97	& --02 26 07.9	& trans. disc, H$\alpha$?		& $<$0.27 & ... 						\\ 
873229	& Haro 5--7			& 05 38 01.07	& --02 45 38.0	& H$\alpha$, II				& $<$0.41 & ... 						\\ 
588270	& S\,Ori J053805.5--023557	& 05 38 05.52	& --02 35 57.1	& II					& $<$0.20 & ... 						\\ 
717307  & [W96] 4771--0950		& 05 38 06.50   & --02 28 49.4  & Li~{\sc i}				& ...     & [FPS2006] 6. In PN gap					\\ 
520267  & S\,Ori 70			& 05 38 10.10   & --02 36 26.0  & low $g$?                              & $<$0.34 & ... \\
757321  & [SWW2004] 233			& 05 38 13.20   & --02 26 08.8  & Li~{\sc i}, H$\alpha$, II		& ...     & [FPS2006] 11. In PN border					\\ 
447254  & S\,Ori J053816.0--023805	& 05 38 16.10   & --02 38 04.9  & Li~{\sc i}, low~$g$			& $<$0.29 & ... 						\\ 
488237	& S\,Ori 27			& 05 38 17.42	& --02 40 24.3	& Li~{\sc i}, H$\alpha$, low~$g$	& $<$0.27 & ... 						\\ 
498234  & S\,Ori J053817.8--024050	& 05 38 17.78   & --02 40 50.1  & II					& ...     & [FPS2006] 17. In PN border. Marginally detected in MOS2	\\ 
396273	& S\,Ori J053818.2--023539	& 05 38 18.35	& --02 35 38.6	& low~$g$				& $<$0.29 & [FPS2006] 19 				\\ 
387252  & S\,Ori J053820.1--023802	& 05 38 20.21   & --02 38 01.6  & Li~{\sc i}, H$\alpha$, II		& $<$0.34 & [FPS2006] 20. Close to PN border. Marginally detected	\\ 
380287	& [W96] r053820--0234		& 05 38 20.50	& --02 34 09.0	& Li~{\sc i}, H$\alpha$			& ...     & ... 						\\ 
379292	& S\,Ori J053821.3--023336	& 05 38 21.38	& --02 33 36.3	& Li~{\sc i}, low~$g$			& ...     & ... 						\\ 
329261	& [SWW2004] 207			& 05 38 23.08	& --02 36 49.4	& low~$g$, II				& ...     & ... 						\\ 
399314	& S\,Ori 18			& 05 38 25.68	& --02 31 21.7	& Li~{\sc i}, low~$g$			& ...     & In MOS gap. Out of PN field			\\ 
265282	& Kiso A--0976 329		& 05 38 27.51	& --02 35 04.2	& Li~{\sc i}, H$\alpha$, II		& ...     & [FPS2006] 33. In MOS border. Out of PN field \\ 
\noalign{\smallskip}
\hline
\end{tabular}
$$
\begin{list}{}{}
\item[$^{a}$] For each star/brown dwarf, the flux upper limit was calculated from its count-rate
              using a Raymond Smith model with $T = 1$\,keV and $N_\mathrm{H} = 0.27
              \times 10^{21}$\,cm$^{-2}$. 
              To determine the count-rate, we used a 15 arcsec extraction radius in 
              the position of the Mayrit source.
\end{list} 
\end{table*}

\begin{figure*}
\centering
\includegraphics[width=1.00\textwidth]{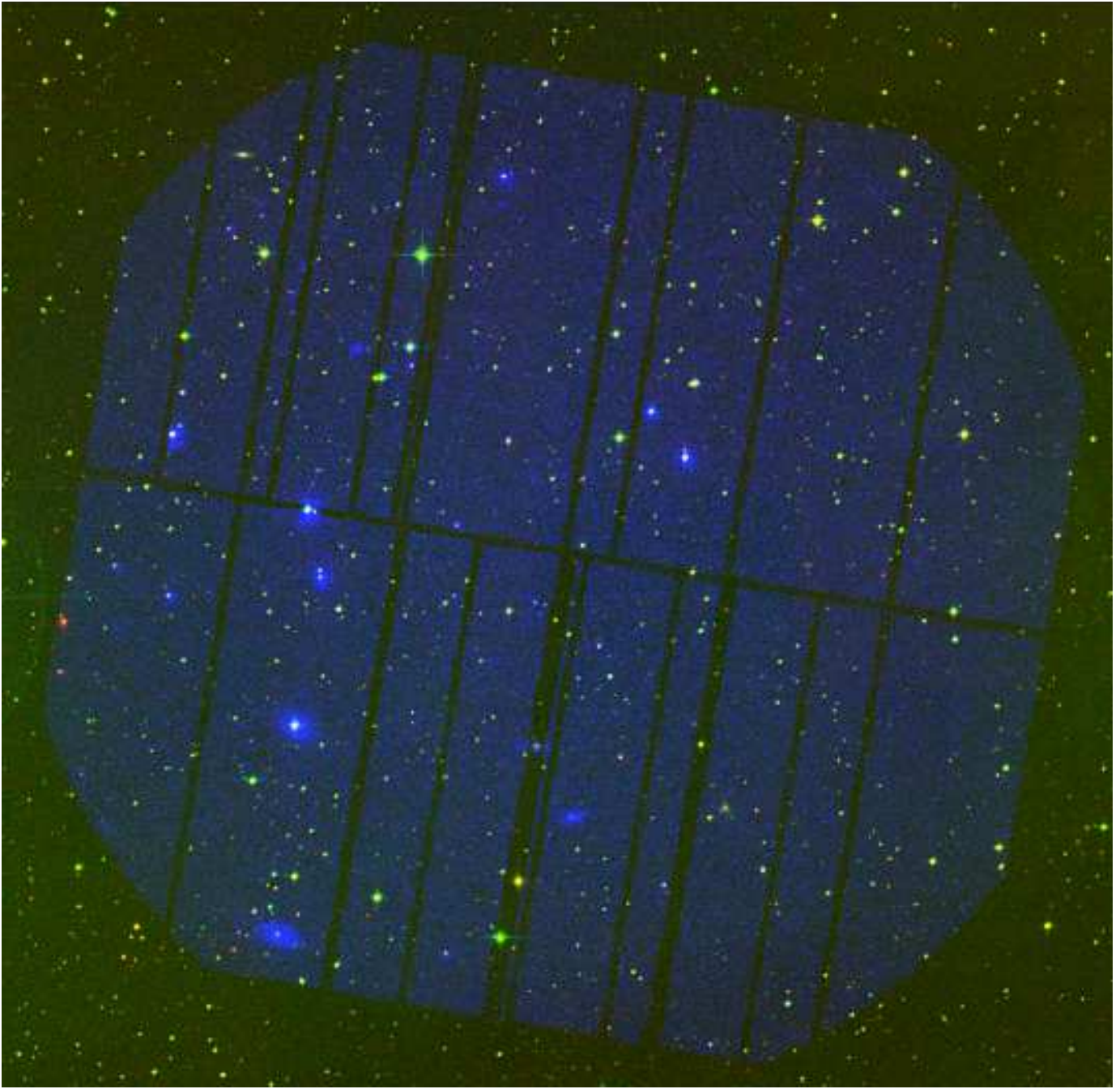}
\caption{False colour image combining 2MASS $K_{\rm s}$ (red), photographic
$R_F$ DSS2 Red (green), and our {\em XMM-Newton} EPIC+MOS1+MOS2 (blue) data.
The field of view of {\em XMM-Newton} (approximate $30\times30$\,arcmin$^2$) 
is clearly discernible. North is up and east is left.}   
\label{falsecolour}
\end{figure*}
%

\begin{figure*}
\centering
\includegraphics[width=0.49\textwidth]{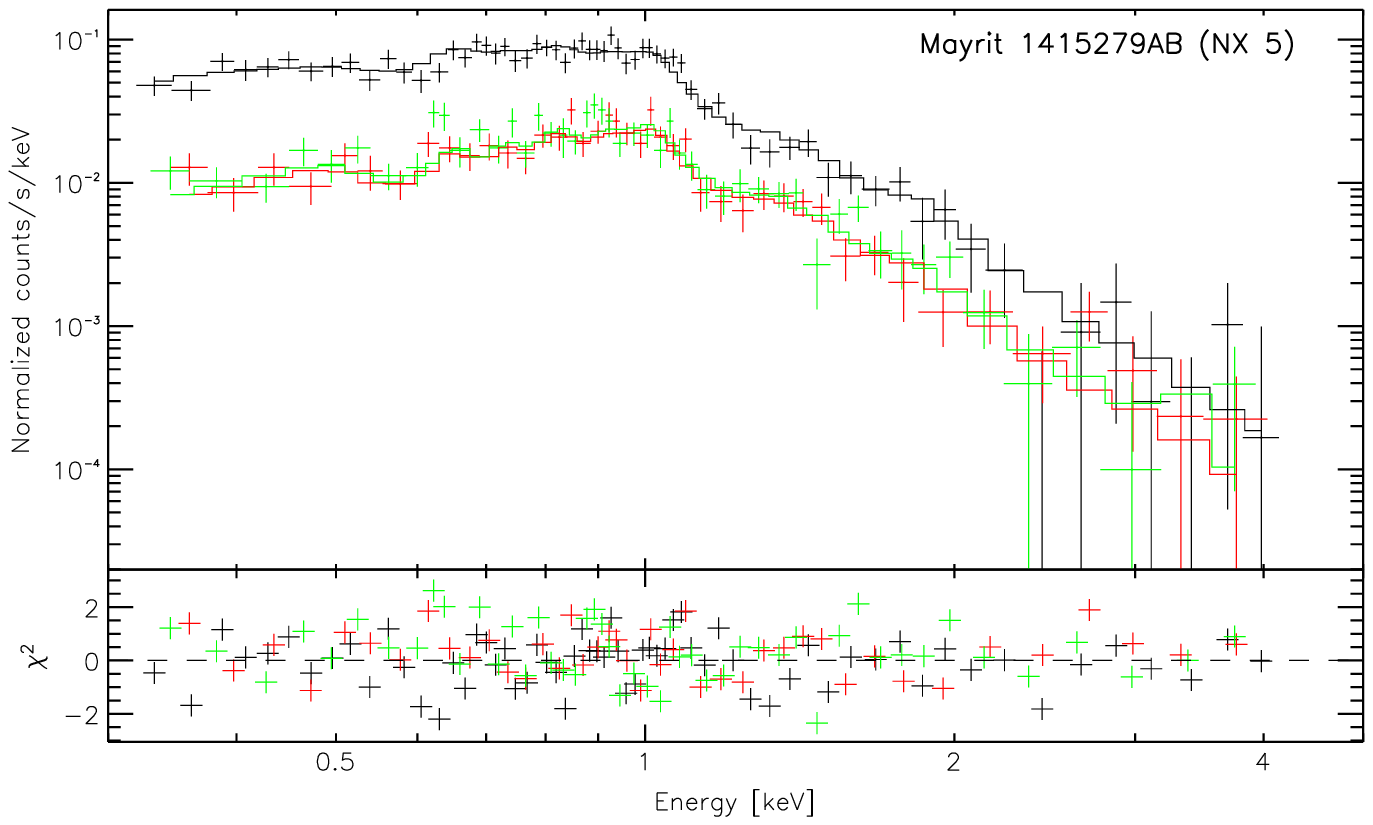}
\includegraphics[width=0.49\textwidth]{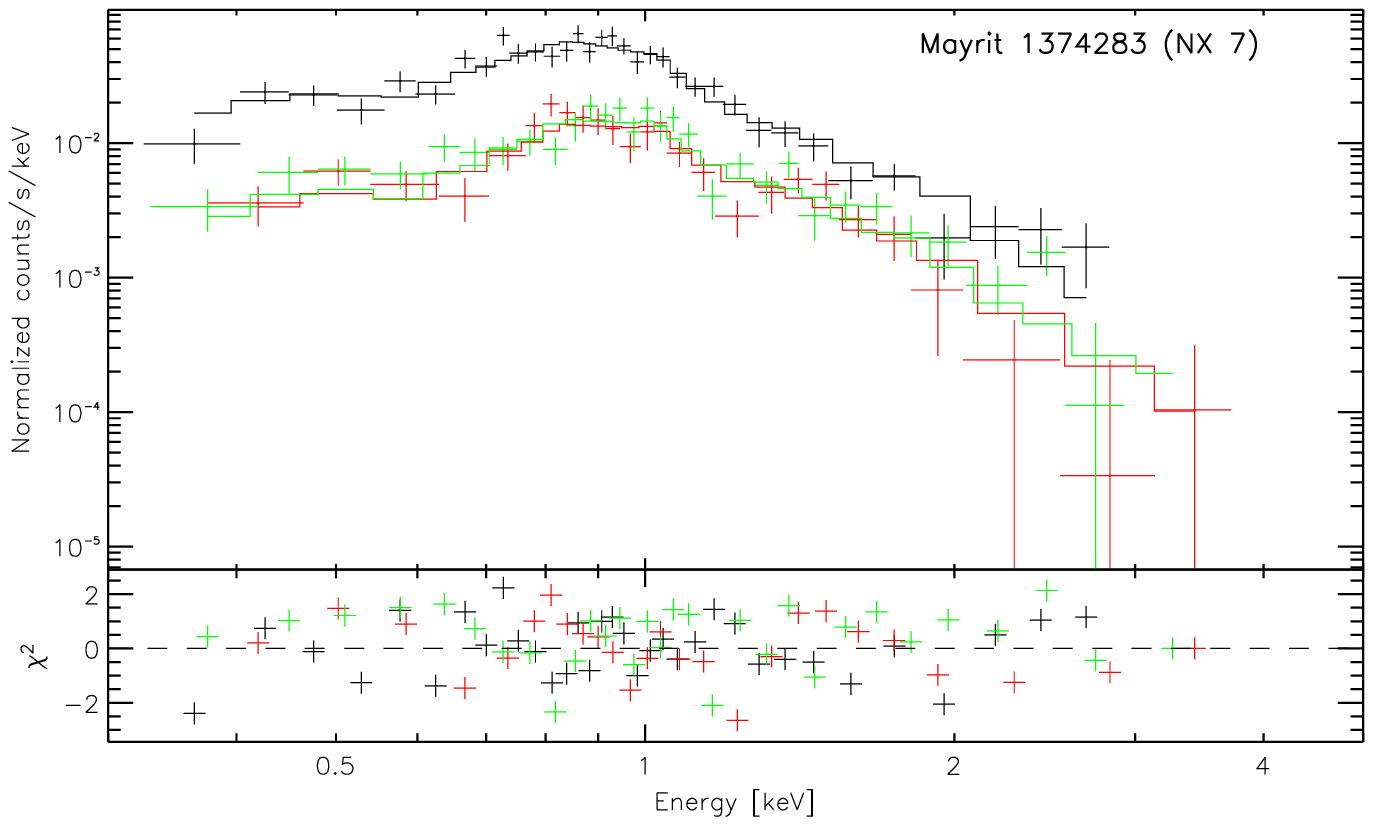}
\includegraphics[width=0.49\textwidth]{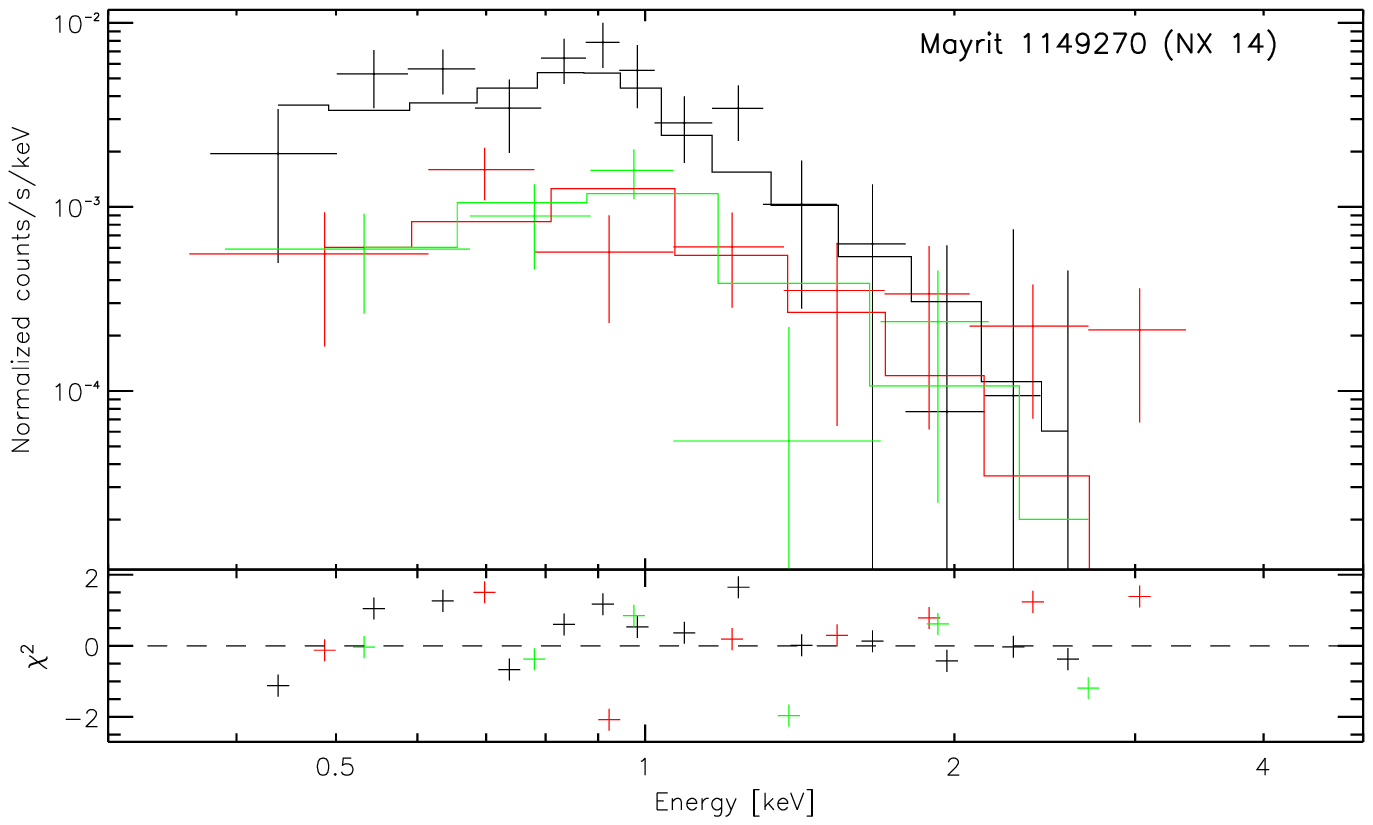}
\includegraphics[width=0.49\textwidth]{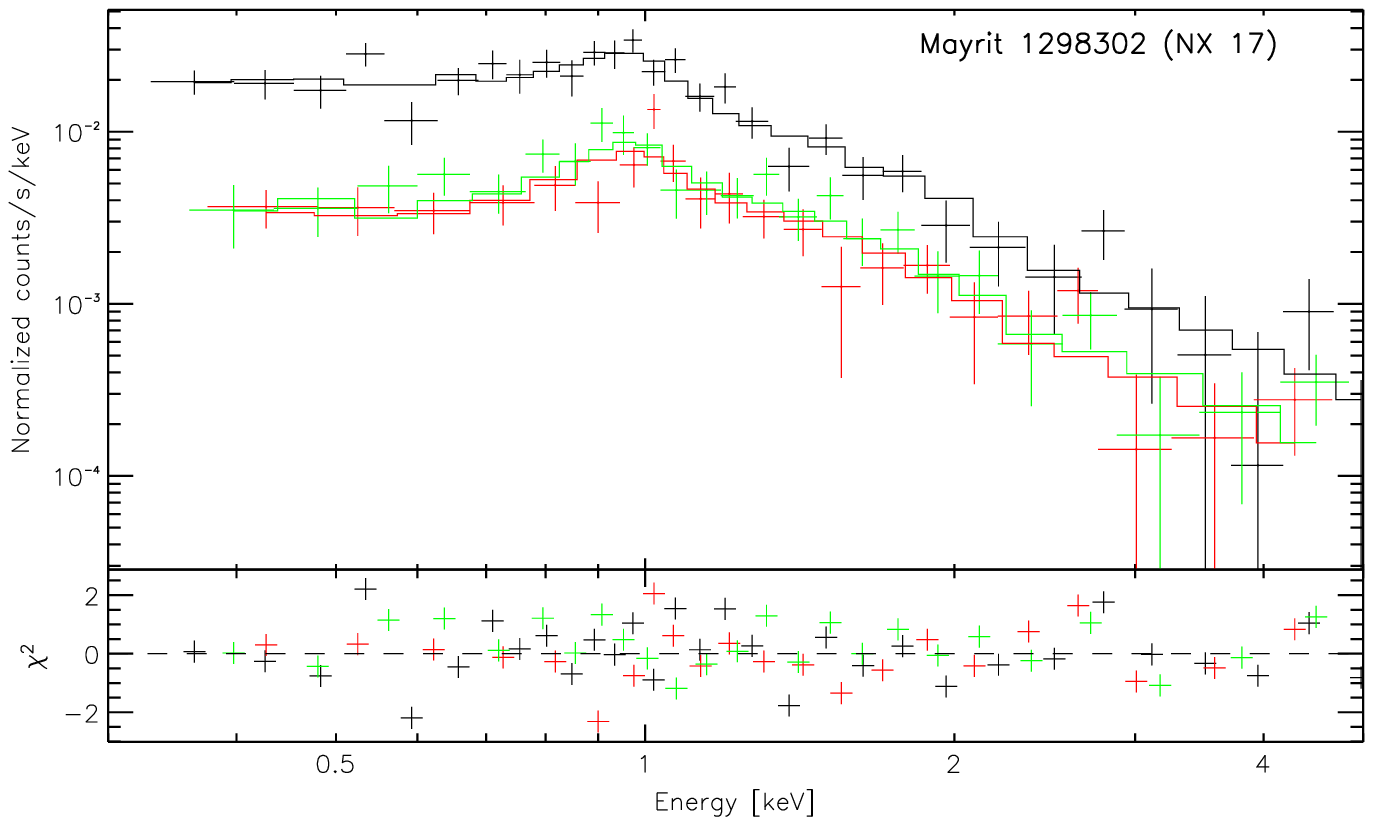}
\includegraphics[width=0.49\textwidth]{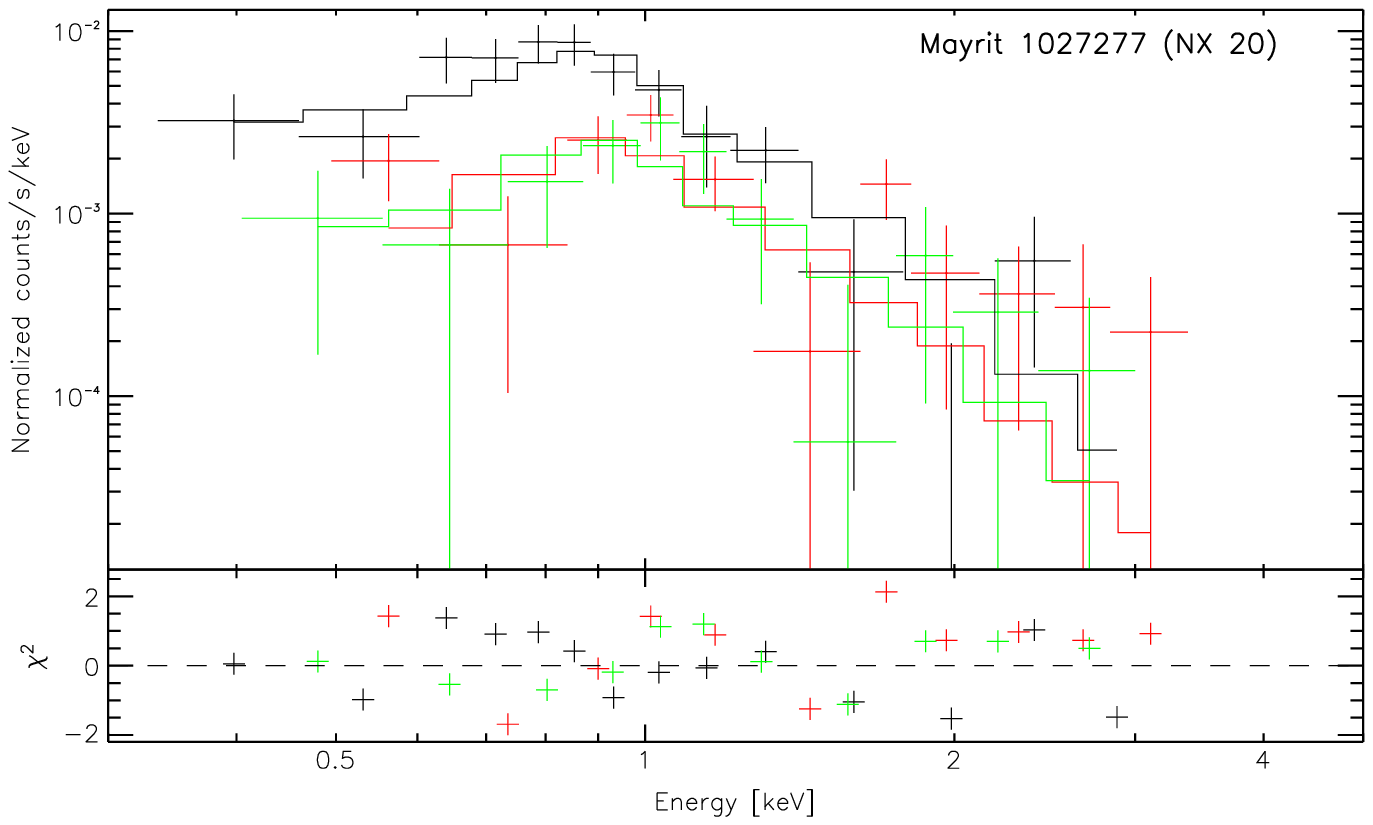}
\includegraphics[width=0.49\textwidth]{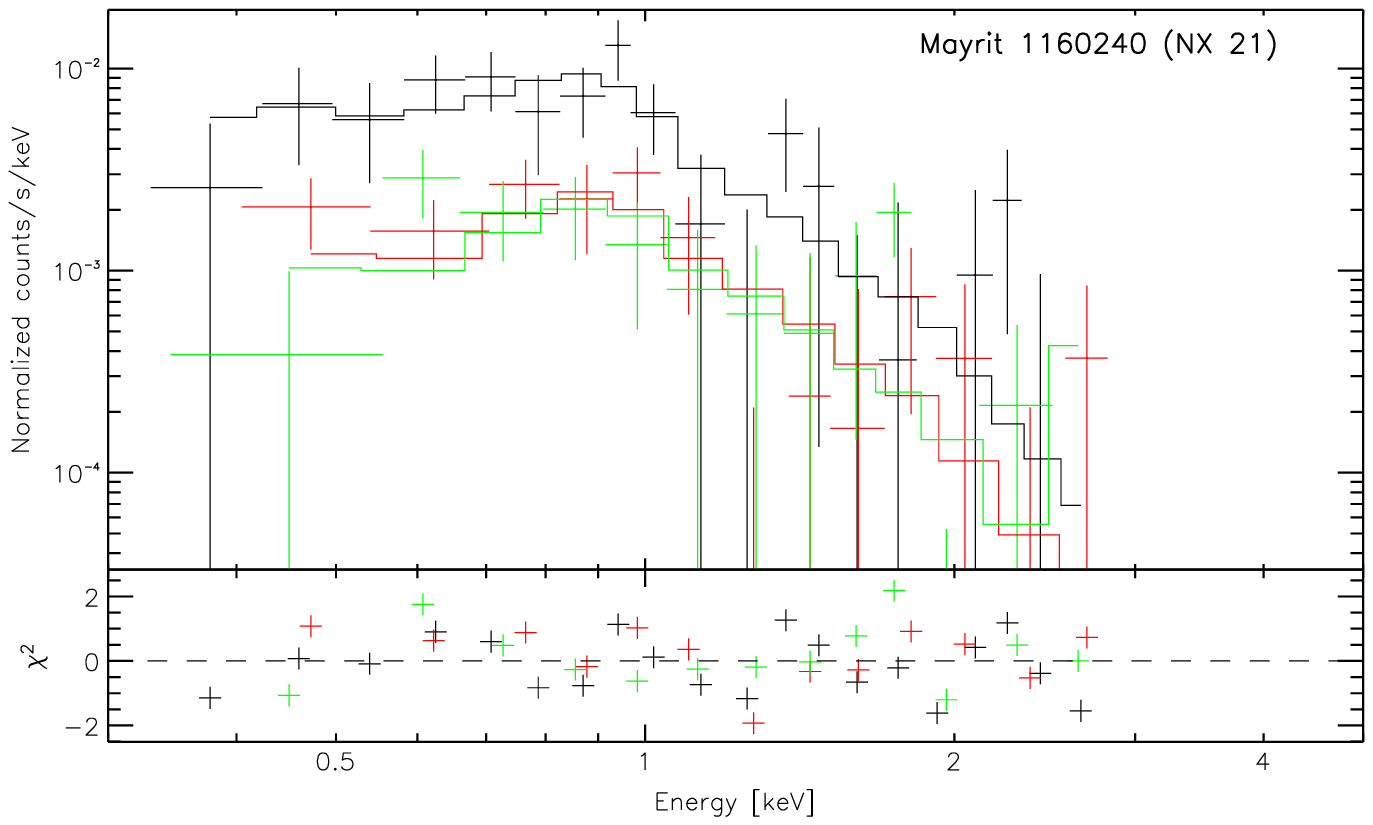}
\caption{Spectral energy distributions of six young stars and candidates in
the $\sigma$~Orionis cluster. Each name is labelled. The spectra were 
binned so that each bin contains a minimum of 10 counts. 
Different line styles correspond to the PN [black], MOS1 [red], and MOS2
[green] cameras and their corresponding spectral fittings 
(see Table\,\ref{table.xray}). The bottom part of each panel illustrates 
the quality of the fitting.}
\label{nx05-21}
\end{figure*}
%

\begin{figure*}
\centering
\includegraphics[width=0.49\textwidth]{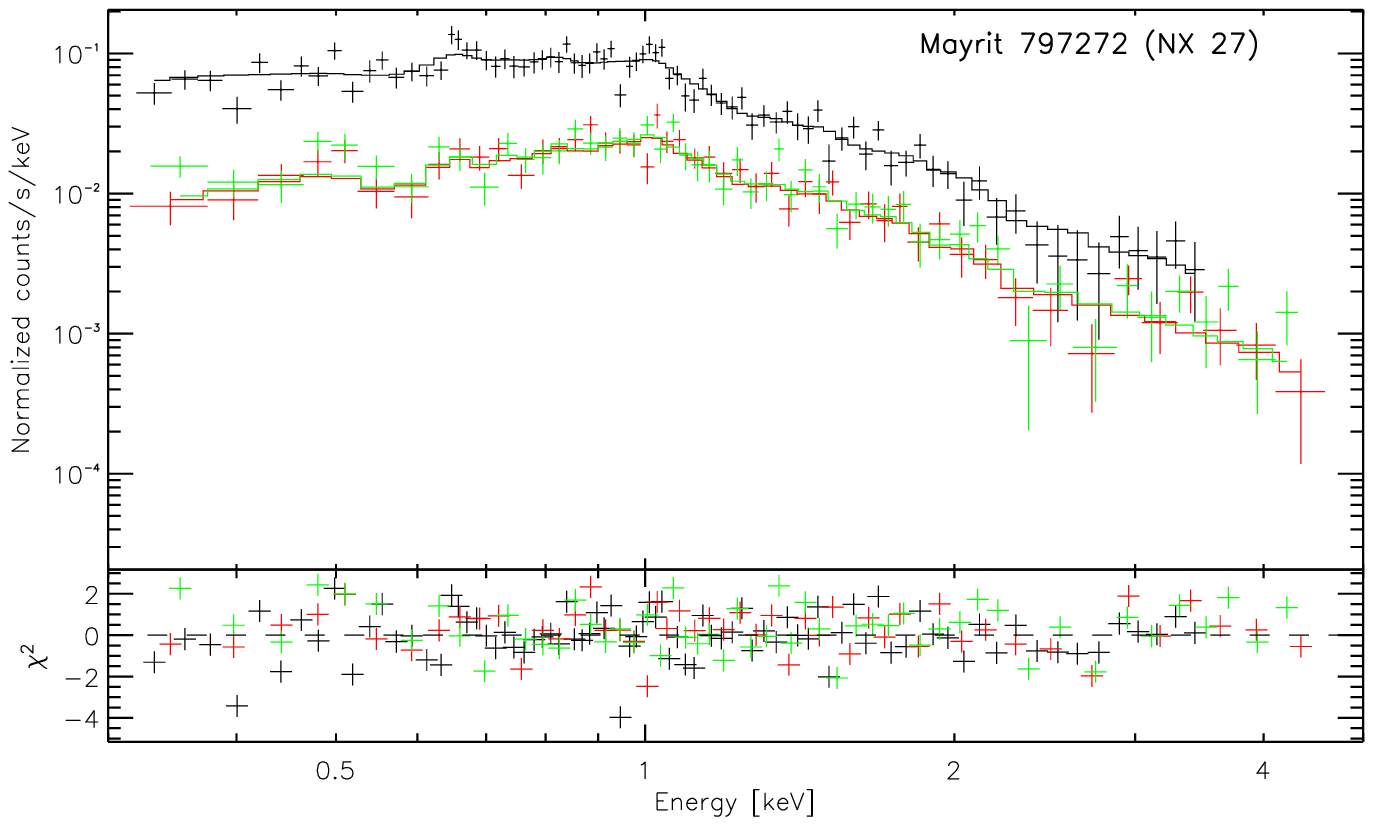}
\includegraphics[width=0.49\textwidth]{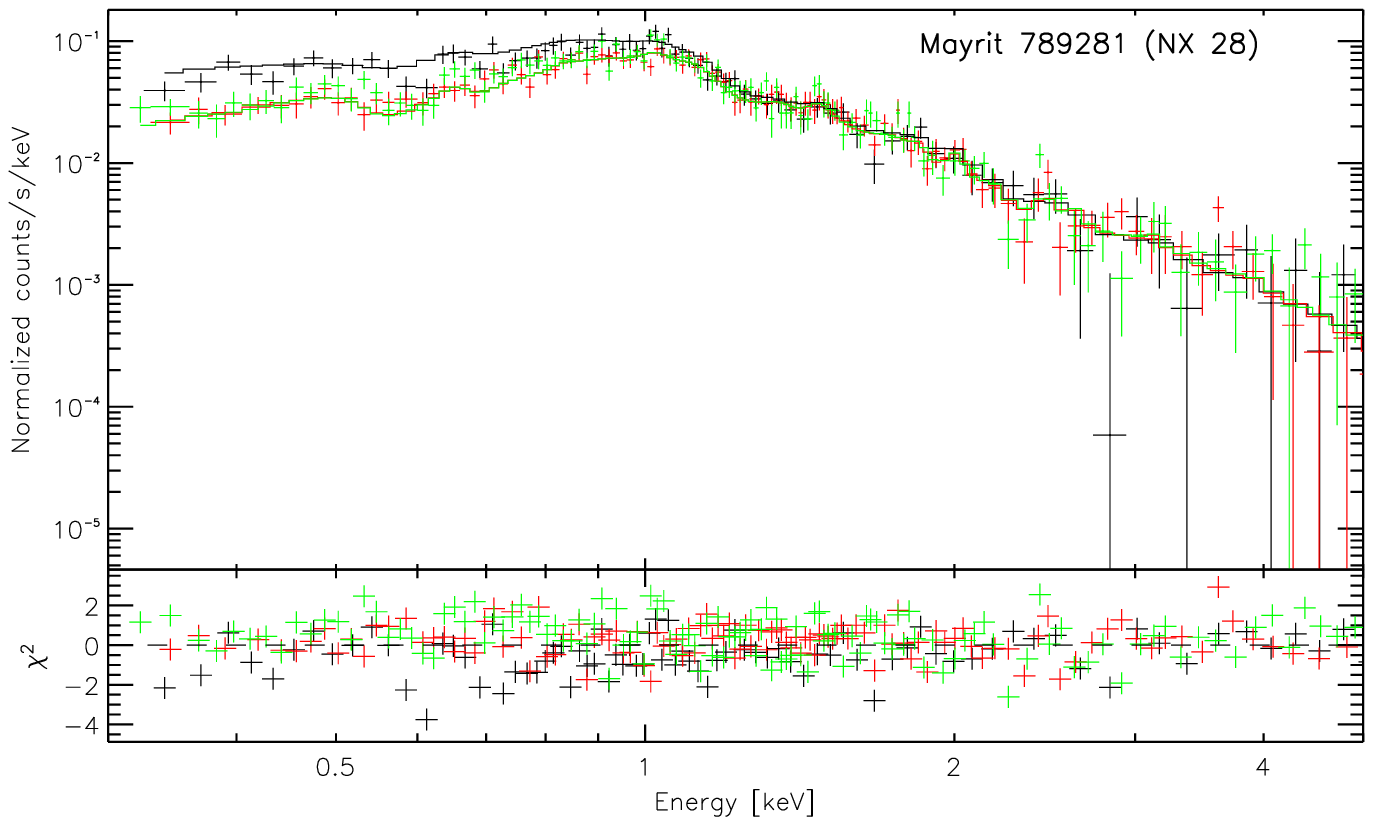}
\includegraphics[width=0.49\textwidth]{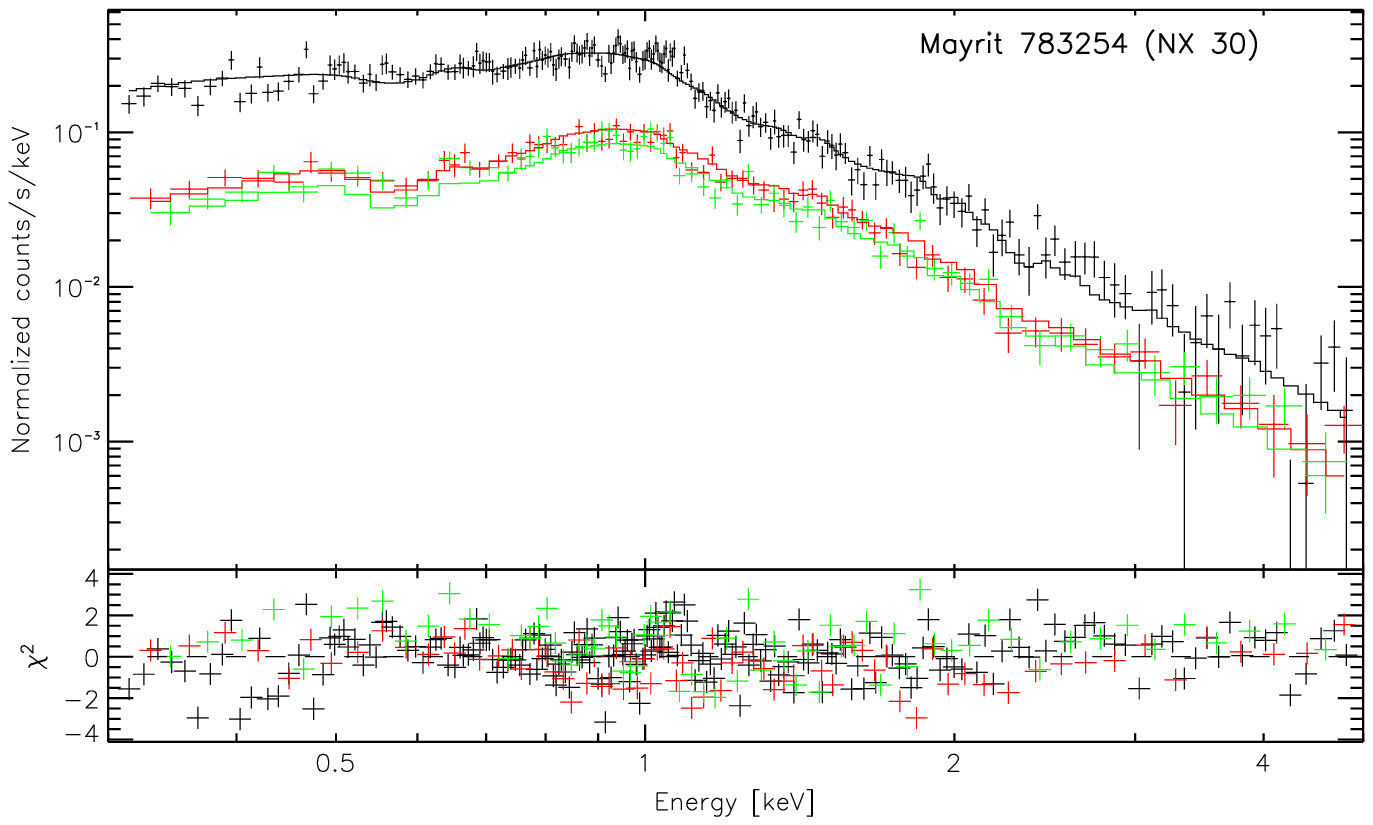}
\includegraphics[width=0.49\textwidth]{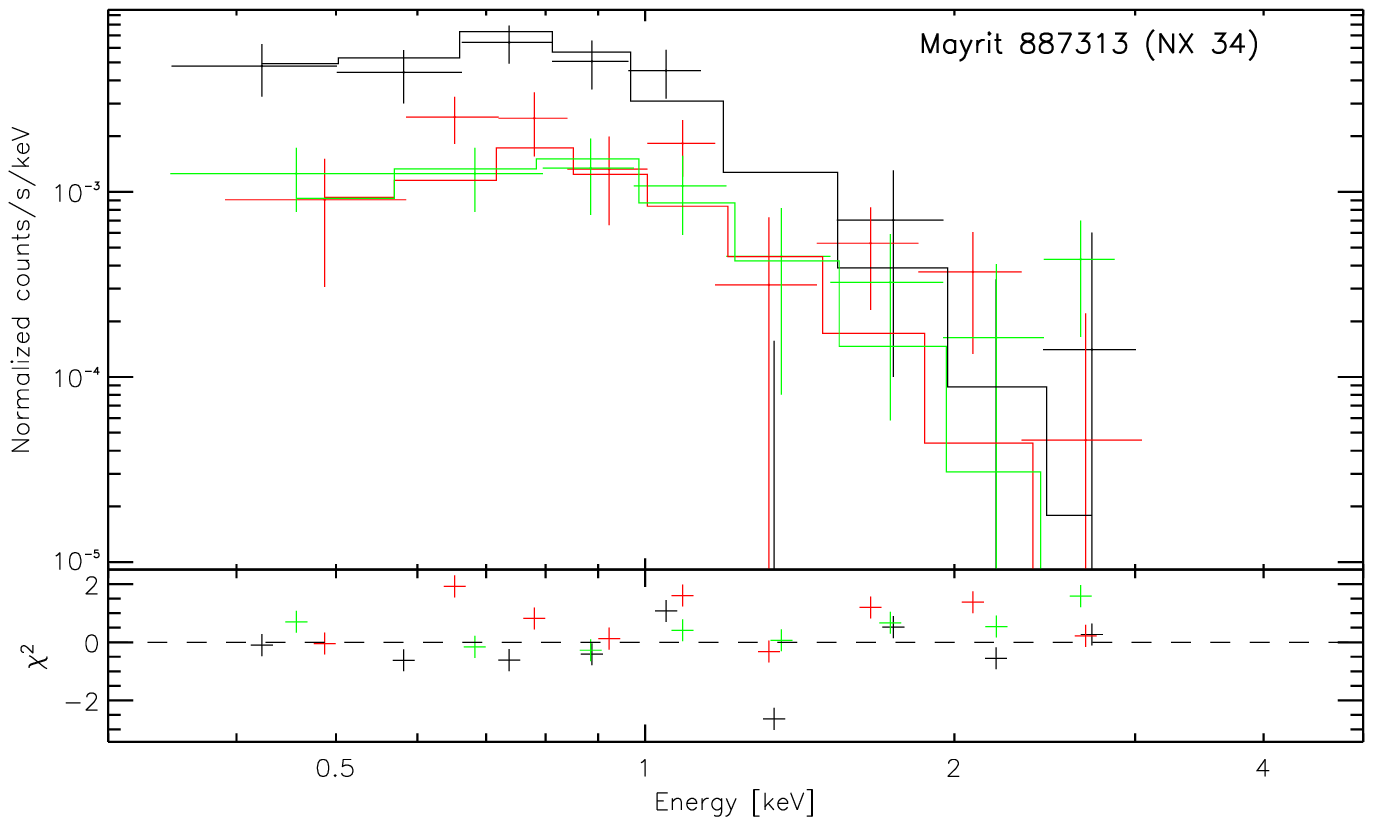}
\includegraphics[width=0.49\textwidth]{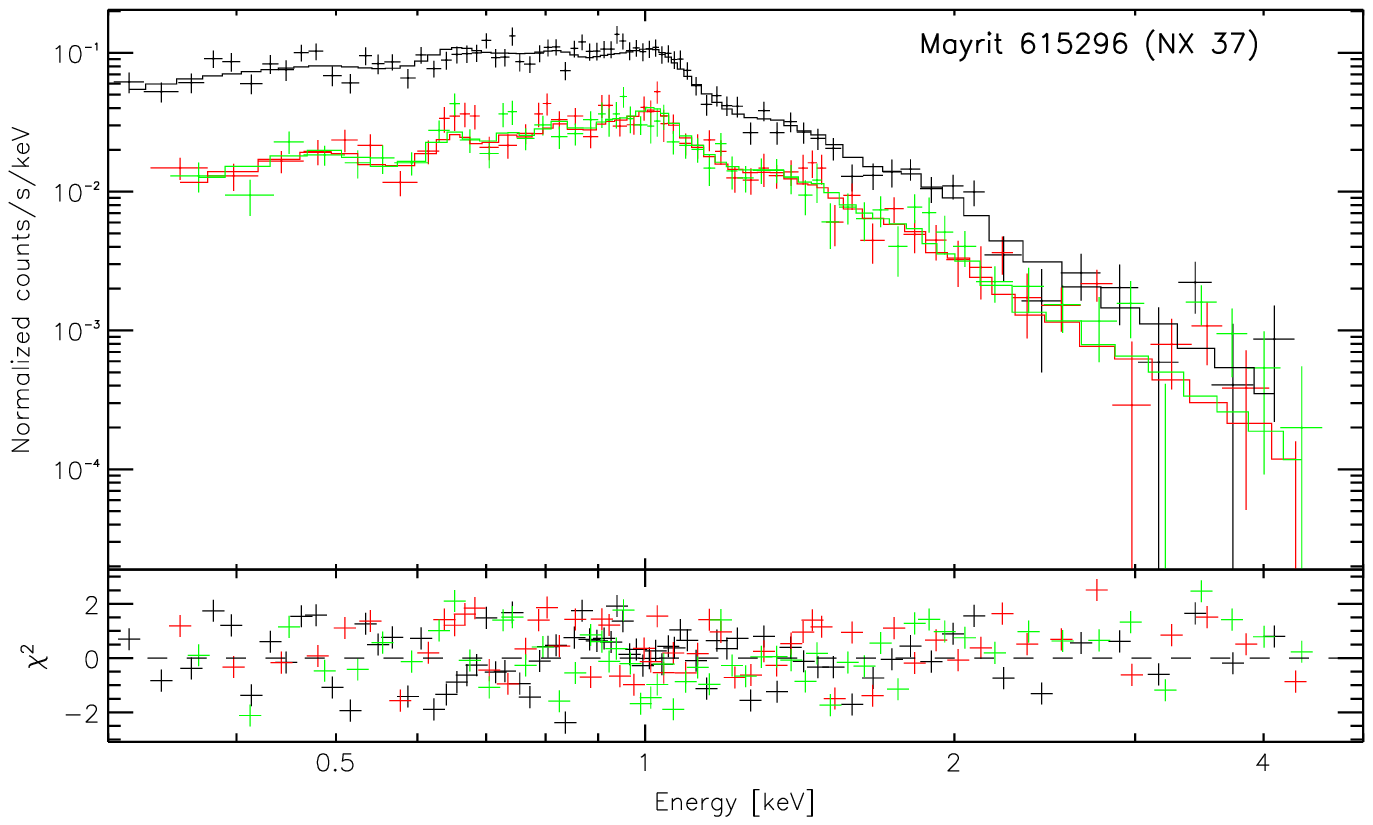}
\includegraphics[width=0.49\textwidth]{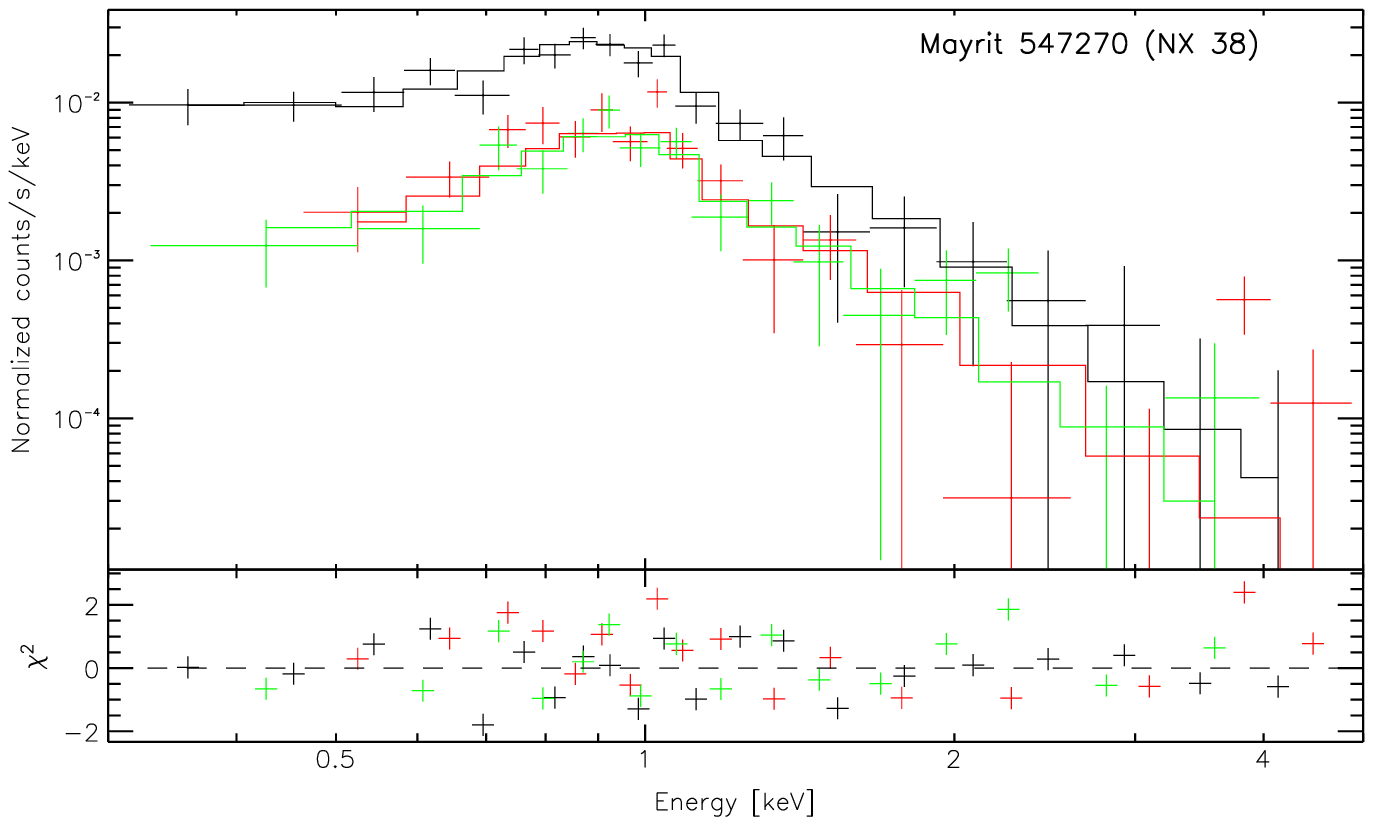}
\caption{Same as Fig.~\ref{nx05-21}, but for the remaining six young stars
and candidates in the $\sigma$~Orionis cluster.
The panel corresponding to NX~27 (top left) is for the $3T$~fitting 
(see Table\,\ref{table.xray}).}
\label{nx27-38}
\end{figure*}
%

\begin{figure*}
\centering
\includegraphics[width=0.49\textwidth]{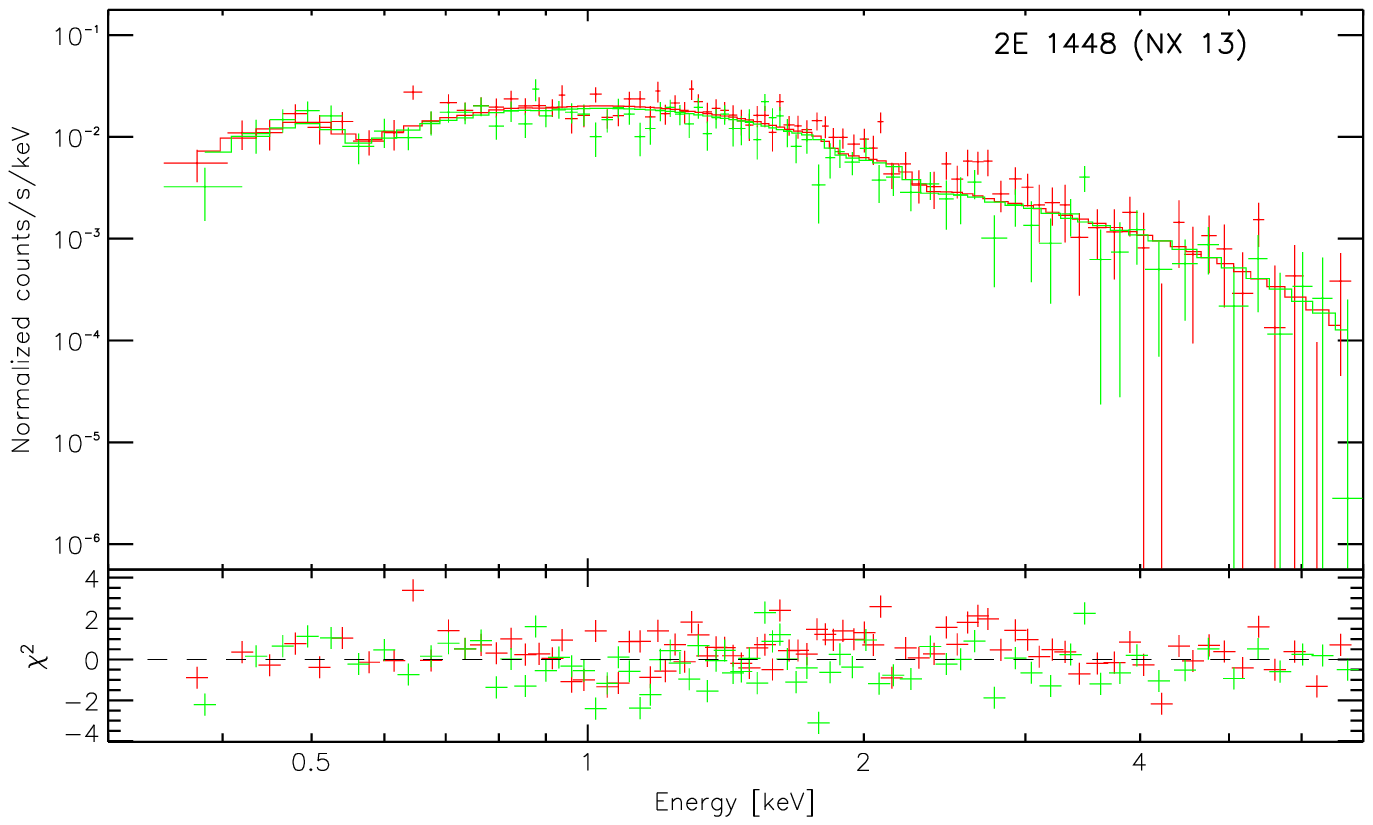}
\includegraphics[width=0.49\textwidth]{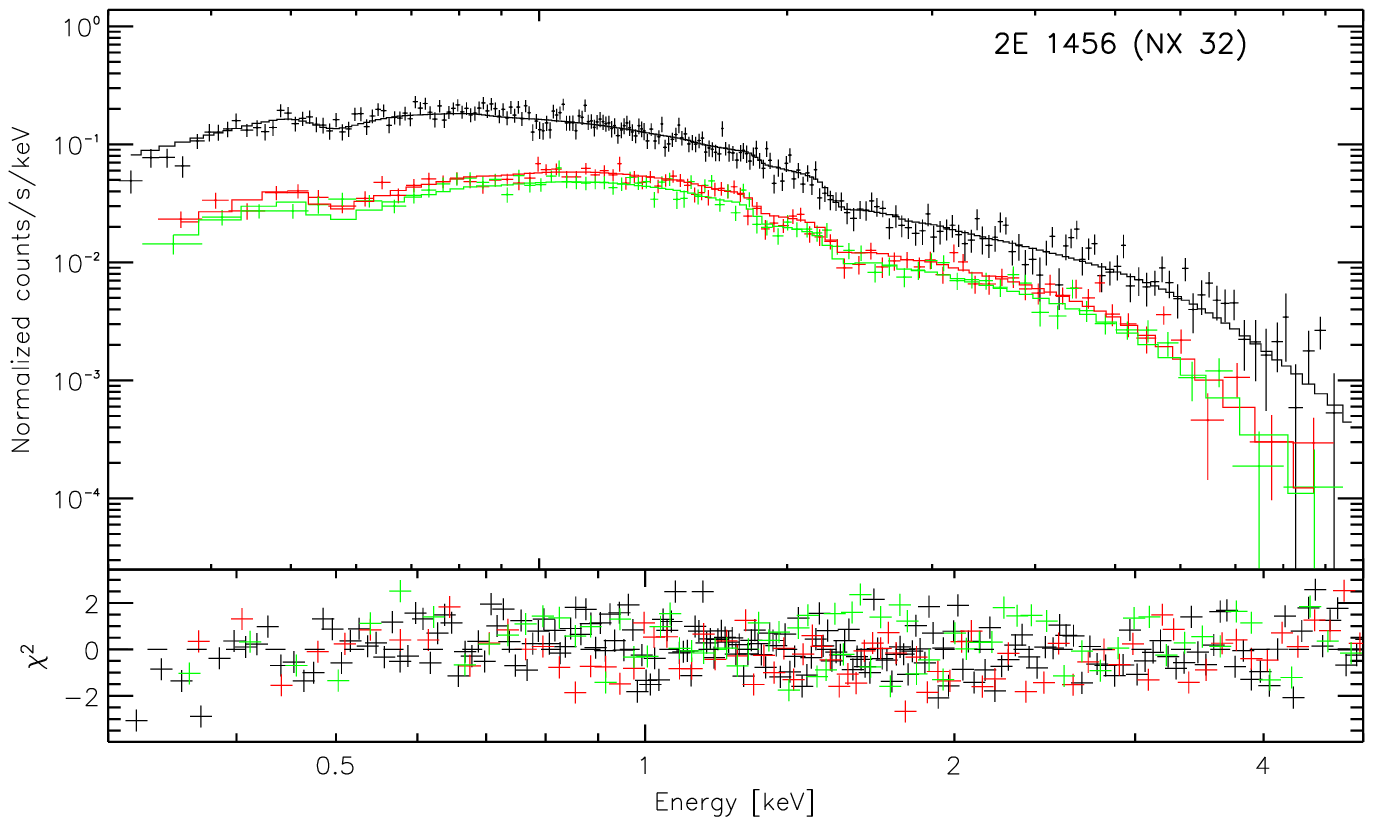}
\caption{Same as Fig.~\ref{nx05-21}, but for two AGNs in 
the field: NX~13 and NX~32. The fittings were made using 
a power-law with parameters $N_\mathrm{H} = 2.05 \times
10^{21}$ cm$^{-2}$, photon power-law index $\Gamma = 2.43$
for NX~13; 
and $N_\mathrm{H} = 1.33 \times 10^{21}$ cm$^{-2}$,  
photon power law index $ \Gamma = 1.82$ for NX~32.} 
\label{fig.agns}
\end{figure*}

\end{document}